\documentclass[lettersize,journal]{IEEEtran}
\usepackage{amssymb}
\usepackage{amsmath}
\usepackage{cite}
\usepackage{url}
\usepackage{xcolor}
\usepackage{cite,graphicx,amsmath,amssymb}
\usepackage{subfigure}
\usepackage{citesort}
\usepackage{fancyhdr}
\usepackage{mdwmath}
\usepackage{mdwtab}
\usepackage{caption}
\usepackage{amsthm}
\usepackage{setspace}
\usepackage{algorithm}
\usepackage{algorithmic}
\usepackage{makecell}
\usepackage{diagbox}
\usepackage{balance}

\newtheorem{remark}{Remark}\newtheorem{theorem}{Theorem}

\newtheorem{lemma}{Lemma}

\newtheorem{corollary}{Corollary}

\allowdisplaybreaks
\makeatletter
\def\ScaleIfNeeded{%
\ifdim\Gin@nat@width>\linewidth \linewidth \else \Gin@nat@width
\fi } \makeatother
\begin{document}
\title{Coexisting Passive RIS and Active Relay Assisted NOMA Systems
}

\author{
Ao~Huang,
 Li~Guo,~\IEEEmembership{Member,~IEEE,}
 Xidong~Mu,~\IEEEmembership{Member,~IEEE,}
 Chao~Dong,~\IEEEmembership{Member,~IEEE,}
  \\and
 Yuanwei~Liu,~\IEEEmembership{Senior Member,~IEEE}

\thanks{Part of this work was presented at the IEEE International Conference on Communications (ICC), Seoul, South Korea, May 16–20, 2022~\cite{ICC_Ao}.}
\thanks{Ao Huang and Li Guo are with the Key Laboratory of Universal Wireless Communications, Ministry of Education, Beijing University of Posts and Telecommunications, Beijing 100876, China, also with the School of Artificial Intelligence, Beijing University of Posts and Telecommunications, Beijing 100876, China, and also with the National Engineering Research Center for Mobile Internet Security Technology, Beijing University of Posts and Telecommunications, Beijing 100876, China (email:\{huangao, guoli\}@bupt.edu.cn).}
\thanks{Xidong Mu and Yuanwei Liu are with the School of Electronic Engineering and Computer Science, Queen Mary University of London, London E1 4NS, U.K. (e-mail: \{xidong.mu, yuanwei.liu\}@qmul.ac.uk).}
\thanks{Chao Dong is with the Key Laboratory of Universal Wireless Communications, Ministry of Education, Beijing University of Posts and Telecommunications, Beijing 100876, China, and also with the School of Artificial Intelligence, Beijing University of Posts and Telecommunications, Beijing 100876, China (email: dongchao@bupt.edu.cn).}
}
\maketitle
\begin{abstract}
A novel coexisting passive reconfigurable intelligent surface (RIS) and active decode-and-forward (DF) relay assisted non-orthogonal multiple access (NOMA) transmission framework is proposed. In particular, two communication protocols are conceived, namely \textit{Hybrid NOMA} (H-NOMA) and \textit{Full NOMA} (F-NOMA). Based on the proposed two protocols, both the sum rate maximization and max-min rate fairness problems are formulated for jointly optimizing the power allocation at the access point and relay as well as the passive beamforming design at the RIS. To tackle the non-convex problems, an alternating optimization (AO) based algorithm is first developed, where the transmit power and the RIS phase-shift are alternatingly optimized by leveraging the  two-dimensional search and rank-relaxed difference-of-convex (DC) programming, respectively. Then, a two-layer penalty based joint optimization (JO) algorithm is developed to jointly optimize the resource allocation coefficients within each iteration. Finally, numerical results demonstrate that: i) the proposed coexisting RIS and relay assisted transmission framework is capable of achieving a significant user performance improvement than conventional schemes without RIS or relay; ii) compared with the AO algorithm, the JO algorithm requires less execution time at the cost of a slight performance loss;  and iii) the H-NOMA and F-NOMA protocols are generally preferable for ensuring user rate fairness and enhancing user sum rate, respectively.
\end{abstract}
\section{Introduction}
With the global promotion of the fifth generation (5G) wireless services and applications, growing research efforts have been devoted to the upcoming beyond 5G (B5G) and sixth-generation (6G) wireless communication networks~\cite{Modulation_Cai}. As a promising paradigm for enhancing the spectrum-efficiency (SE) and energy-efficiency (EE) in wireless communications, reconfigurable intelligent surfaces (RISs) have received extensive attention from both academia and industry~\cite{Towards_Wu,Vision_6G}. An RIS is a planar array composed of massive tunable unit elements, each of them can passively reflect the incident electromagnetic wave while changing its amplitude and phase-shift. By appropriately reconfiguring the electromagnetic response of RIS elements, the desired signals can be enhanced and the undesired interfering signals can be mitigated at the destination with the aid of the a so-called ‘Smart Radio Environment’~\cite{Smart_Renzo}. Due to the nearly passive working mode, RISs only passively reflect signals with no radio frequency (RF) chains, thus reducing hardware costs and energy consumption~\cite{Practical_Abeywickrama}. Moreover, RISs can be flexibly deployed on building facades, road signs, lamp posts, etc. Given the above advantages, RISs are regarded as a promising technology for next-generation wireless networks. 

On the other hand, non-orthogonal multiple access (NOMA), which has been emerged as one of the vital enabling multiple access techniques to support future Internet-of-Everything (IoE) and Mobile Internet~\cite{NOMA_Proc,NGMA}. The key idea of NOMA is to allow multiple users to occupy the same resource block (i.e. time, frequency and code), where superposition coding (SC) and successive interference cancellation (SIC) are employed at the transmitter and receiver, respectively~\cite{NOMA_Proc}. 
As a result, to cater for the explosive growth of mobile data traffic, NOMA is a strong candidate technology for the next-generation wireless networks, expected to serve massive numbers of users in a more resource-efficient manner. Compared with conventional orthogonal transmission strategies, significant performance enhancement can be achieved by NOMA, including higher system SE and EE~\cite{Performance_Zhiguo,Cooperative_Yunawei}, and better user fairness~\cite{Exploiting_Xidong}.
\vspace{-0.3cm}
\subsection{Prior Works}
\subsubsection{Studies on RIS-aided Systems}
The potential benefits of deploying RISs in wireless communication systems have been exploited in many prior works.
For instance, Huang \textit{et al.}~\cite{Huang_EE} proposed an energy consumption model for the RIS reflection unit. On this basis, joint beamforming design is investigated to maximize EE in the RIS-assisted multiple-input single-output (MISO) system while guaranteeing individual link budget for each user. Wu \textit{et al.}~\cite{Spectrum_Sensing} proposed a novel RIS-enhanced energy detection scheme for spectrum sensing to cope with the case of severe channel fading. Yu \textit{et al.}~\cite{MISO_Xianghao} studied the power-efficient resource allocation design for RIS-assisted multiuser MISO system with the goal of minimizing the transmit power at the access point (AP). Pan \textit{et al.}~\cite{MIMO_Cunhua} revealed that RISs can assist the signal interference mitigation of cell edge users. The weighted user sum rate of the RIS-assisted multi-cell multiple-input multiple-output (MIMO) multi-user communication system is maximized by the joint beamforming design. Hua \textit{et al.}~\cite{Multipoint_Meng} proposed to deploy a RIS to assist the joint processing coordinated multipoint (CoMP) transmission from multiple base stations to multiple cell-edge users.
In addition, Yildirim \textit{et al.}~\cite{Hybrid_Yildirim} introduced two hybrid transmission schemes combining RIS with active relay to assist transmission for coverage extension. Moreover, Zheng and Zhang~\cite{IRS_Relaying} unlocked the potential of RIS controller in relaying information, by jointly optimizing the time allocation and the RIS passive beamforming design to maximize the achievable rate of the proposed system. Very recently, to overcome the fundamental limitation of “double fading” effect, a new concept of \textit{active} RIS has been proposed, this more general RIS design can achieve a significantly higher transmission rate compared with those via passive RIS~\cite{Active_Zhang,Active_Dongfang}.
\subsubsection{Studies on RIS-NOMA Systems}
To further enhance multiple access capabilities, growing research efforts have been devoted to investigating the integration of RIS and NOMA technologies.
Specifically, Ding \textit{et al.}~\cite{Simple_design} proposed a novel RIS-assisted NOMA communication model to maximize the total number of served users by deploying RISs in cell edge areas. Based on this setup, the outage performance under an on-off RIS control scheme is analyzed. Zheng \textit{et al.}~\cite{User_Pairing} compared the theoretical performance of RIS-assisted NOMA and orthogonal multiple access (OMA), showing that NOMA and OMA prefer asymmetric and symmetric user pairing schemes, respectively. To characterize the Pareto boundary of the capacity and rate regions for both multiple access strategies, Mu \textit{et al.}~\cite{Capacity_Xidong} proposed to jointly optimize the RIS reflection matrix and wireless resource allocation under the constraints of discrete phase shifts and a finite number of RIS reconfiguration times. Zhu \textit{et al.}~\cite{RIS_NOMA_Jianyue} considered an RIS-assisted NOMA network to minimize the total transmit power under the fundamental two-user scenario, where the RIS-assisted zero-forcing beamforming (ZFBF) design is introduced for comparison. In another aspect, Chen \textit{et al.}~\cite{MEC} conceived a RIS-aided wireless powered mobile edge computing (MEC) system, in which both time division multiple access (TDMA) and NOMA schemes are considered for MEC uplink offloading. Mu \textit{et al.}~\cite{Deployment_Xidong} investigated the joint RIS deployment and multiple access design for downlink RIS-assisted multi-user networks, and concluded the optimal deployment principle of RISs in different multiple access communications with the goal of maximizing the weighted sum rate.
\subsection{Motivations and Contributions}
The working principle of RISs is similar to the full-duplex (FD) relays, but provides a nearly passive mode of operation without imposing self-interference issues~\cite{loop_interference}. However, with the natural passive architecture, RISs cannot amplify the magnitude of the incident signals limiting the achievable performance gain. Considering the stringent communication requirements in future wireless networks, we need to break through the dilemma of technological conflict~\cite{Wu2019IRS,Relaying_Bjornson}. Instead of relying only on a single technology, we have to strike the good convergence of passive RISs and active relays. The coexistence of these two technologies would lead to the following benefits. On the one hand, for the conventional active relay-assisted transmission, passive RISs can enhance the transmission quality in a seamless and low-cost manner. On the other hand, for the pure RIS-assisted transmission, active relays provide additional signal processing capabilities for facilitating sophisticated transmission strategies. Given the aforementioned benefits of converging RISs and relays, some initial works~\cite{Hybrid_Yildirim,IRS_Relaying} have studied the coexisting RIS and relay assisted communication designs in point-to-point scenarios. However, to the best of the authors' knowledge, the communication protocol and design for the coexisting RIS and relay assisted multi-user systems have not been studied, yet. This provides the main motivations for this work.

Against the above background, in this work, we propose a novel coexisting passive RIS and active relay assisted NOMA transmission framework, where one AP communicates with two NOMA users with the aid of both one RIS and one decode-and-forward (DF) relay. The information transmission consists of two stages, namely the information broadcasting stage (IB-stage) and the information relaying stage (IR-stage). For facilitating the transmission in the proposed framework, two communication protocols are conceived, namely the \textit{hybrid NOMA} (H-NOMA) and the \textit{full NOMA} (F-NOMA). The main difference between the two protocols is that only the NOMA weak user's information is relayed in H-NOMA while both users' information is relayed employing NOMA in F-NOMA. For each protocol, we investigate the communication design for maximizing the sum rate and guaranteeing the user rate fairness. The main contributions of this paper can be summarized as follows:
\begin{itemize}
	\item 
	 We propose a novel coexisting passive RIS and active relay assisted downlink NOMA transmission framework, in which RIS and DF relay are deployed for assisting the communication between the AP and a pair of NOMA users. Two communication protocols are proposed, namely H-NOMA and F-NOMA. 
	 We evaluate the performance of each proposed protocol under two optimization criteria, including the sum rate maximization and maximin fairness, by jointly optimizing the power allocation at the AP and relay as well as the RIS passive beamforming design.
	 
	\item 
	We first propose an alternating optimization (AO) based algorithm, where the original problem is decomposed into two subproblems. For the power allocation optimization subproblem, the optimal solution is selected by exhaustively two-dimensional searching among all possible candidates. For the RIS phase-shift configuration subproblem, a difference-of-convex (DC) based rank-one relaxation is leveraged to obtain the stationary point solution of the passive beamforming design.
	\item 
	We further propose a joint optimization (JO) algorithm to jointly optimize the resource allocation, in which the power allocation coefficient and the RIS phase-shift coefficient can be updated synchronously after each iterative procedure. In particular, we first perform a series of transformations to resolve the coupling between optimization variables, thereby facilitating the development of the two-layer iterative algorithm based on successive convex approximation (SCA) technique.	
	 \item 
	Our numerical results unveil that: 1) our proposed coexisting RIS and relay assisted downlink NOMA transmission framework is capable of achieving promising performance improvement, compared with the conventional communications without RIS or relay; 2) the convergence speed of the JO algorithm is much faster than that of the AO algorithm, while the latter algorithm achieves a higher performance; 3) H-NOMA is capable of ensuring the user rate fairness, while F-NOMA is superior for improving the user sum rate.
\end{itemize}
\subsection{Organization and Notation}
The rest of this paper is organized as follows. Section II presents the system model and two communication protocols, based on which the optimization problem for designing the coexisting RIS and DF relay assisted NOMA system are formulated. In Section III and Section IV, we propose the AO and JO algorithms to solve the original optimization problems, respectively. Numerical results are provided in Section V to verify the effectiveness of the proposed designs compared to baseline schemes, which is followed by the conclusions in Section VI.\\
\indent \emph{Notations:} Scalars, vectors, and matrices are denoted by lower-case, bold-face lower-case, and bold-face upper-case letters, respectively. ${\mathbb{C}^{N \times 1}}$ denotes the space of $N \times 1$ complex-valued vectors. ${{\mathbf{a}}^H}$ and $\left\| {\mathbf{a}} \right\|$ denote the conjugate transpose and the Euclidean norm of vector ${\mathbf{a}}$, respectively. ${\textup {diag}}\mathbf{(a)}$ denotes a diagonal matrix with the elements of vector ${\mathbf{a}}$ on the main diagonal. The distribution of a circularly symmetric complex Gaussian (CSCG) random variable with mean $\mu $ and variance ${\sigma ^2}$ is denoted by ${\mathcal{CN}}\left( {\mu,\sigma ^2} \right)$. ${\textup{Rank}}\mathbf{(A)}$ and ${\textup {Tr}}\mathbf{(A)} $ denote the rank and the trace of matrix $\mathbf{A}$, respectively. ${\textup {diag}}\mathbf{(A)} $ denotes a vector whose elements are extracted from the main diagonal elements of matrix $\mathbf{A}$. ${{\mathbf{A}}} \succeq 0$ indicates that $\mathbf{A}$ is a positive semidefinite matrix. ${\left\| {\mathbf{A}} \right\|_*}$ and ${\left\| {\mathbf{A}} \right\|_2}$ denote the nuclear norm and spectral norm, respectively.
\section{System Model and Problem Formulation}
\begin{figure*}[t]
	\centering	
	\setlength{\belowcaptionskip}{+0.2cm}   %调整图片标题与下文距离
	\subfigure[]{
		\includegraphics[width=3.41in]{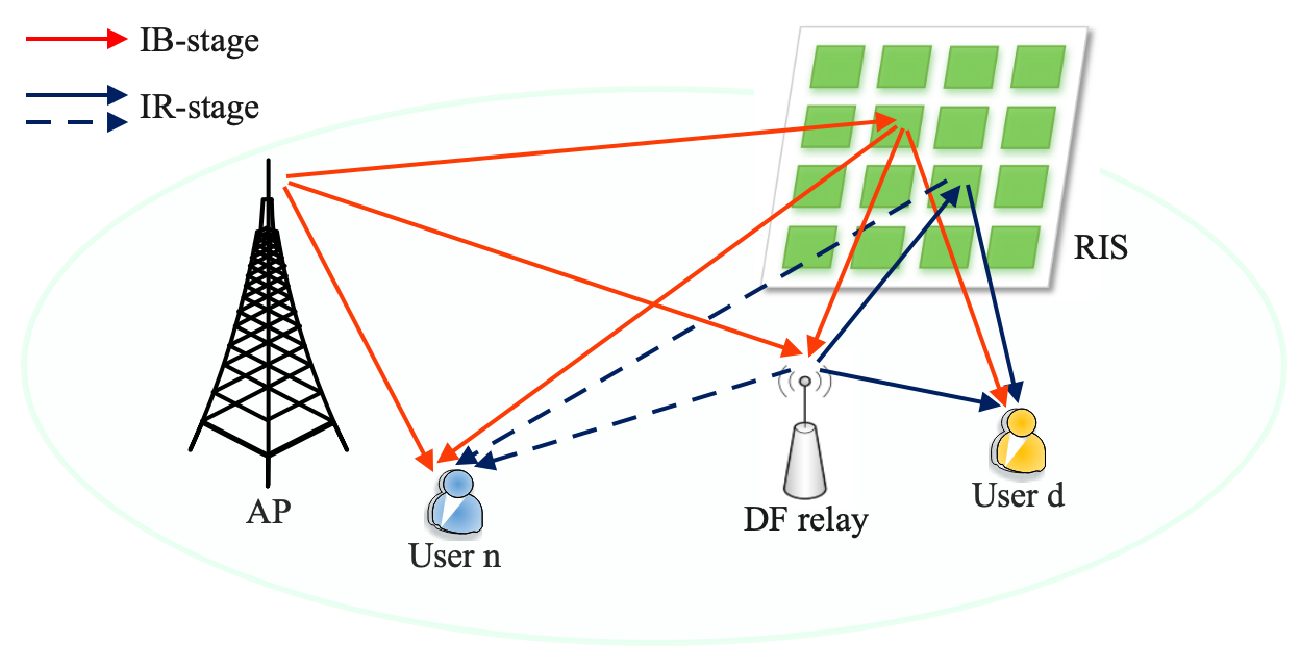}
	}
	\;
	\subfigure[]{
		\includegraphics[width=3.41in]{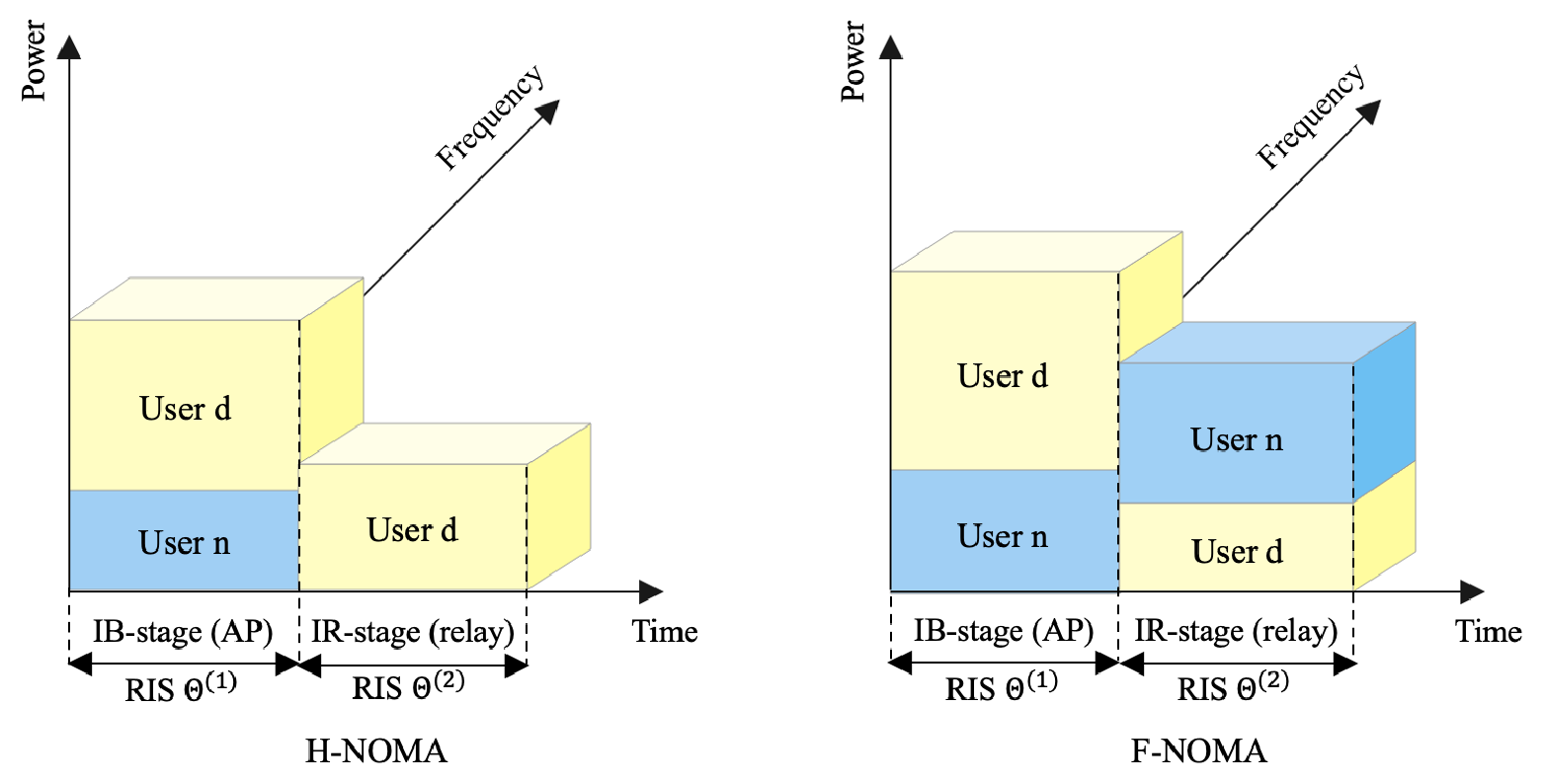}
	}
	\caption{Illustration of the considered coexisting RIS and relay assisted NOMA transmission framework and resource allocation strategies. (a) System model design. (b) Resource allocation based on each communication protocol.}
	\label{System model}
\end{figure*}
In this section, we present the system model for the considered coexisting RIS and relay assisted NOMA transmission framework, based on this, two communication protocols are proposed and the problem formulations are developed.
\subsection{System Model}
As illustrated in Fig. 1(a), we consider a downlink coexisting passive RIS and active relay assisted NOMA transmission framework, where a RIS and a DF relay are deployed to assist communication from an AP to a pair of users. The system is assumed to be of narrow band at a given carrier frequency, and the carrier frequency $f_c$ and the system bandwidth $B$ satisfy $B\ll f_c$~\cite{Tutorial}. To reveal the fundamental performance gain and design insights for the practical system, it is assumed that the AP and both users are equipped with a single antenna. The RIS is composed of $M$ elements. For ease of exposition, we refer to the nearby user of the AP as user $n$, and the distant user as user $d$. Considering a more practical scenario for user distribution, there exists a direct link $r_{{\rm A}n}\in\mathbb{C}^{1 \times 1}$ to establish communication from the AP to user $n$. Whereas the link between the AP and user $d$ is assumed to be severely blocked, due to e.g., the occlusion of the buildings and infrastructures, which is a very challenging scenario for conventional wireless communication systems$\footnote{To address this issue, properly deploying RIS and relay to assist the transmission in the communication dead zone is an appealing solution. Moreover, the solutions proposed in this work are also applicable for the case where the direct links between the AP and both users exist.}$. $ r_{{\rm AR}}\in\mathbb{C}^{1 \times 1}$ and $ r_{{\rm R}k}\in\mathbb{C}^{1 \times 1}$ denote the channels between form the AP to relay and the relay to user $k\in\mathcal{K}=\{n,d\}$, respectively. The RIS-assisted links, including the channels from the AP to RIS, the relay to RIS, the RIS to relay, and the RIS to user $k$ are denoted by
$\mathbf{G}_{ {\rm AI} }\in\mathbb{C}^{M \times 1}$, $\mathbf{G}_{{\rm RI} }\in\mathbb{C}^{M \times 1}$, $\mathbf{h}_{\rm{IR} }\in\mathbb{C}^{M \times 1}$ and $\mathbf{h}_{{\rm I}k }\in\mathbb{C}^{M \times 1}$, respectively. The transmission model for the two-user setup consists of two stages:
	\begin{itemize}
		\item 
		\textbf{IB-stage:} 
		The AP sends the superposition coding to serve the two users. Meanwhile, the DF relay also receives the same signal.
		\item 
		\textbf{IR-stage :} 
		According to the adopted communication protocol, the decoded message is forwarded from the DF relay to the designated users.
	\end{itemize}
\noindent
In this paper, we assume that perfect channel state information (CSI) is available at the AP to facilitate the resource allocation design. The perfect CSI for the direct links without RIS can be obtained by applying the conventional channel estimation methods~\cite{CoMP_Wenhao}. However, the acquisition of accurate CSI for RIS-assisted channels is a challenging task due to the near-passive operation mode of RISs. To address this issue,  the one of the existing channel estimation techniques can be adopted to efficiently obtain the perfect CSI, e.g., brute-force method~\cite{OFDM_Yifei}, compressive-sensing based method~\cite{Hang_CSI}, and deep learning based method~\cite{Channel_Estimation_Chang}.
\subsection{Communication Protocols}
In this paper, two communication protocols are proposed. As shown in Fig. 1(a), dashed lines represent the signals that may occur during the IR-stage. The DF relay can switch between two protocols, choosing to serve a single distant user or to serve both users simultaneously. This has led to the discussions on whether NOMA continues to be needed at the IR-stage. The following is a detailed description of the two protocols, and the resource allocation strategy based on each protocol is illustrated in Fig. 1(b).
\begin{itemize}
	\item 
	\textbf{\textit{Hybrid NOMA} Transmission:} 
	For H-NOMA, during the IB-stage, the AP transmits signals to paired users utilizing NOMA technology. 
	While, during the IR-stage, the DF relay is selected to assist distant user communication based on OMA technology.
	\item 
	\textbf{\textit{Full NOMA} Transmission:} 
    For F-NOMA, using NOMA technology, data services for paired users allocated in the same resource block can be simultaneously supported at both stages. Note that, the roles of the original nearby and distant users during the IR-stage are reversed with respect to the transmitter due to channel reassignment.
\end{itemize}
\subsection{Signal Model}
 \subsubsection{\textup{\textbf{IB-stage}}}
 The AP simultaneously transmits the signals of two users by adopting the superposition coding, the transmitted signal based on NOMA protocol is given by
 \begin{equation}
 	 x^{{(1)}}=\sqrt{\alpha_{n} P_{a}} \textit s_{n}+\sqrt{\alpha_{d} P_{a}} s_{d},
 \end{equation}
 where $s_{n}$ and $s_{d}$ are the normalized signal for user $n$ and user $d$, respectively, such that $\mathbb{E}\{s_{ n}^2\}=\mathbb{E}\{s_{d}^2\}=1$. $P_{a}$ denotes transmit power budget at the AP. $\alpha_{n}$ and $\alpha_{d}$ are the corresponding power allocation coefficients, with $0\le \alpha_{n}\le \alpha_{d}$ and $\alpha_{n}+\alpha_{d}=1$. The observation at user $ n$ is
 \begin{equation}
 	y_{n}^{(1)}= \left( r_{{\rm A}n}+ \mathbf{h}_{{\rm I}n}^H\mathbf{ \Theta }^{(1)}\mathbf{G}_{{\rm AI}}\right) x^{(1)}+{z}_{n}^{(1)}, 
 \end{equation}
where $ z_{n}^{(1)}\sim \mathcal{CN} (0, \sigma_{n}^2)$ denotes the IB-stage additive white Gaussian noise (AWGN) at user ${n}$ with zero mean and variance $\sigma_{n}^2$. $\mathbf{\Theta}^{(1)}={\textup{diag}}\{e^{j\theta_{1}^{(1)}},e^{j\theta_{2}^{(1)}},\cdots,e^{j\theta_{M}^{(1)}}\}$ denotes the IB-stage reflection coefficients matrix of the RIS, where $ \theta_{m}^{(1)}\in[0,2\pi)$ is the phase-shift coefficient of the $m$-th reflecting element, which is assumed to be continuously adjustable to investigate the maximum system performance. According to NOMA protocol, user ${n}$ first employs SIC$\footnote{Note that, with appropriate modifications, the algorithms developed in this work are also applicable to the case where the inter-user interference cannot be completely eliminated by employing SIC~\cite{Unified_Xinwei}.}$ to decode the signal ${s}_{d}$ of user $d$ having a higher power allocation coefficient, the received signal-to-interferenceplus-noise ratio (SINR) is given by
 \begin{equation}
 	\textup{SINR}_{n{\rightarrow}d}^{(1)}=\frac{\alpha_{d} P_{a}| r_{{\rm A}n} +  \mathbf{h}_{{\rm I}n}^H\mathbf{ \Theta }^{(1)}\mathbf{G}_{{\rm AI} }|^2}{\alpha_{n} P_{a}| r_{{\rm A}n } +  \mathbf{h}_{{\rm I}n}^H\mathbf{ \Theta }^{(1)}\mathbf{G}_{{\rm AI} }|^2+\sigma_{ n}^2}.
 \end{equation}
 After removing the signal ${s}_{d}$, user $n$ can decode its own signal $s_{n}$ from the resulting reception, the received SINR is given by
  \begin{equation}
 		\textup{SINR}_{n{\rightarrow} n}^{(1)}
 		=\frac{\alpha_{n} P_{a}| r_{{\rm A}n} + \mathbf{h}_{{\rm I}n}^H\mathbf{ \Theta }^{(1)}\mathbf{G}_{{\rm AI} }|^2}{\sigma_{n}^2}.
 \end{equation}
	The observation at user $d$ is
 \begin{equation}
 	y_{d}^{(1)}= \left(\mathbf{h}_{{\rm I}d}^H\mathbf{ \Theta }^{(1)}\mathbf{G}_{{\rm AI} }\right) x^{(1)}+ z_{d}^{(1)},
 \end{equation}
\noindent
 where $z_{d}^{(1)}\sim \mathcal{CN} (0, \sigma_{d}^2)$ is the IB-stage AWGN at user $d$. The received SINR to decode $s_{d}$ is given by
 \begin{equation}
 	\textup{SINR}_{d{\rightarrow}d}^{(1)}=\frac{\alpha_{d} P_{a}| \mathbf{h}_{{\rm I}d}^H\mathbf{ \Theta }^{(1)}\mathbf{G}_{{\rm AI} }|^2}{\alpha_{n} P_{a}| \mathbf{h}_{{\rm I}d}^H\mathbf{ \Theta }^{(1)}\mathbf{G}_{{\rm AI} }|^2+\sigma_{d}^2}.
 \end{equation}
The observation at the DF relay is
 \begin{equation}
 	y_{r}^{(1)}= \left( r_{{\rm AR} } + \mathbf{h}_{\rm{IR}}^H\mathbf{ \Theta }^{(1)}\mathbf{G}_{{\rm AI} }\right) x^{(1)}+ z_{r}^{(1)},
 \end{equation}
 where $z_{r}^{(1)}\sim \mathcal{CN} (0, \sigma_{r}^2)$ is the IB-stage AWGN at DF relay.
 The received SINR to decode $s_{d}$ is given by
 \begin{equation}
 		\textup{SINR}_{r{\rightarrow}d}^{{(1)}}=\frac{\alpha_{d}P_{a}| r_{{\rm AR} } + \mathbf{h}_{\rm{IR}}^H\mathbf{ \Theta }^{(1)}\mathbf{G}_{{\rm AI} }|^2}{\alpha_{n} P_{a}| r_{{\rm AR} } + \mathbf{h}_{\rm{IR}}^H\mathbf{ \Theta }^{(1)}\mathbf{G}_{{\rm AI} }|^2+\sigma_{r}^2},
 \end{equation}
where $\textit R_{r{\rightarrow}d}^{(1)}
=\frac{1}{2}\log_2\left(1+\textup{SINR}_{r{\rightarrow} d}^{{(1)}}\right)$  is the achievable data rate. If the F-NOMA protocol is employed, the relay will decode the signal $s_{n}$ after successfully extracting the signal $s_{d}$. Otherwise, it is not required. The received SINR to decode $s_{n}$ is given by
 \begin{equation}
 		\textup{SINR}_{r{\rightarrow}n}^{(1)}
 		=\frac{\alpha_{n} P_{a}| r_{{\rm AR} } + \mathbf{h}_{\rm{IR}}^H\mathbf{ \Theta }^{(1)}\mathbf{G}_{{\rm AI} }|^2}{\sigma_{r}^2},
 \end{equation}
where $	\textit R_{r{\rightarrow}n}^{(1)}=\frac{1}{2}\log_2\left(1+ \textup{SINR}_{r{\rightarrow}n}^{{(1)}}\right)$ is the achievable data rate.

Assuming that the relay is capable of decoding the two NOMA user’s information, i.e, satisfying the following conditions, 1) $\textit R_{r{\rightarrow}n}^{(1)}\ge\textit R_{n}^{\min}$; and 2) $\textit R_{r{\rightarrow}d}^{(1)}\ge\textit R_{d}^{\min}$, where the $\textit R_{k}^{\min}$ is the target rate for user $k\in\{n, d\}$.
\subsubsection{\textup{\textbf{IR-stage}}}
Based on the adopted communication protocol at the DF relay, the decoded message will be forwarded to the designated users. 
\paragraph{H-NOMA} 
In this case, IR-stage is to achieve the communication between the DF relay and user $d$. As a result, the transmit signal intended to user $d$ is given by
\begin{equation}
	x^{\rm H{(2)}}=\sqrt{ P_{r}}s_{d},
\end{equation}
where $ P_{r}$ denotes the average allowed transmit power at the relay.
The observation at user $d$ is
\begin{equation}
	y^{\rm H{{(2)}}}_{d}= \left(r_{{\rm R}d} + \mathbf{h}_{{\rm I}d}^H\mathbf{ \Theta }^{(2)}\mathbf{G}_{{\rm RI} }\right) x^{\rm H{(2)}}+ z_{d}^{(2)},
\end{equation}
where $z_{d}^{(2)}\sim \mathcal{CN} (0, \sigma_{d}^2)$ is the IR-stage AWGN at user $d$. 
$\mathbf{ \Theta }^{(2)}={\textup{diag}}\{e^{j\theta_{1}^{(2)}},e^{j\theta_{2}^{(2)}},\cdots,e^{j\theta_{M}^{(2)}}\}$
denotes the  IR-stage reflection coefficients matrix of the RIS, where for all $m\in\mathcal{M}$, $\theta_{m}^{(2)}\in[0,2\pi)$ is the phase-shift of the $m$-th reflecting element. Thus, the received SINR at user $d$ is given by
\begin{equation}
	\textup{SINR}_{d{\rightarrow}d}^{\rm H{(2)}}=\frac{ P_{r}| r_{{\rm R}d} +\mathbf{h}_{{\rm I}d}^H\mathbf{ \Theta }^{(2)}\mathbf{G}_{{\rm RI} }|^2}{\sigma_{d}^2}.
\end{equation}
	Maximal ratio combining (MRC) can be utilized to merge the signals from the AP and the DF relay based on the assumption that both signals can be fully resolved at user $d$. Hence, the received SINR at user $d$ to detect its own message after MRC combination is given by~\cite{Cooperative_Yunawei,Hybrid_Gang}
	\begin{equation}
		\tilde{\gamma}^{\rm H}_{{d},\textup{MRC}}=\textup{SINR}_{d{\rightarrow}d}^{{(1)}}+\textup{SINR}_{d{\rightarrow}d}^{\rm H(2)},
	\end{equation}
where $\textit R^{\rm H}_{{d},\textup{MRC}}=\frac{1}{2}\log_2\left(1+\tilde{\gamma}^{\rm H}_{{d},\textup{MRC}}\right)$ is the data rate of user $d$ after MRC.
\noindent However, the rate $\textit R^{\rm H}_{{d},\textup{MRC}}$ can be achieved if and only if the relay has the ability to decode $s_{d}$. Thus, the data rate achieved at user $d$ is bounded by the data rate of relay to decode $s_{d}$, i.e. $\textit R_{r{\rightarrow} d}^{{(1)}}$~\cite{RIS_Elhattab}. Therefore, the final achievable data rate of user $d$ to decode its own signal can be given as
\begin{equation}
	\textit R^{\rm H}_{d}=\min\left\{\textit R_{r{\rightarrow}d}^{{(1)}},\textit R^{\rm H}_{{d},\textup{MRC}}\right\}.
\end{equation}
As a result, the sum rate of the downlink transmission system employing H-NOMA protocol can be expressed as
\begin{equation}
	\textit R^{\rm H}_{sum}=\textit R^{\rm H}_{n}+\textit R^{\rm H}_{d},
\end{equation}
where $\textit R^{\rm H}_{n}=\frac{1}{2}\log_2\left(1+\textup{SINR}_{n{\rightarrow}n}^{(1)}\right)$ is the achievable IB-stage data rate to decode $s_{n}$ at user $n$, indicating that the near user performance is only determined by the communication quality at the IB-stage for H-NOMA.
\paragraph{F-NOMA} 
In this case, to adapt to the new channel link environment, proper power allocation at the DF relay is required.
The superimposed mixture of the signals intended to user $n$ and $d$ at the relay according to the NOMA principle is expressed as
\begin{equation}
	x^{\rm F{(2)}}=\sqrt{\beta_{n} P_{r}} s_{n}+\sqrt{\beta_{d} P_{ r}} s_{d},
\end{equation}
where $\beta_{n}$ and $\beta_{d}$ are the updated power allocation coefficients,  with $0\le\beta_{d}\le \beta_{n}$ and $\beta_{n}+\beta_{d}=1$. The observation at user $d$ is
\begin{equation}
	y^{\rm F{(2)}}_{d}= \left(r_{{\rm R}d} + \mathbf{h}_{{\rm I}d}^H\mathbf{ \Theta }^{(2)}\mathbf{G}_{{\rm RI} }\right) x^{\rm F{(2)}}+ z_{d}^{(2)}.
\end{equation}
Based on NOMA protocol, user $d$ first decodes the signal $s_{ n}$, the received SINR is given by
\begin{equation}
		\textup{SINR}_{d{\rightarrow} n}^{\rm F{(2)}}=\frac{\beta_{n} P_{r}| r_{{\rm R}d} + \mathbf{h}_{{\rm I}d}^H\mathbf{ \Theta }^{(2)}\mathbf{G}_{{\rm RI} }|^2}{\beta_{d} P_{r}| r_{{\rm R}d} + \mathbf{h}_{{\rm I}d}^H\mathbf{ \Theta }^{(2)}\mathbf{G}_{{\rm RI} }|^2+\sigma_{d}^2}.
\end{equation}	
Then, the user can decode its own siganl $s_{d}$, the received SINR is given by
\begin{equation}	
		\textup{SINR}_{d{\rightarrow}d}^{\rm F{(2)}}=\frac{\beta_{d} P_{r}| r_{{\rm R}d} + \mathbf{h}_{{\rm I}d}^H\mathbf{ \Theta }^{(2)}\mathbf{G}_{{\rm RI} }|^2}{\sigma_{d}^2}.
\end{equation}
The observation at user $n$ is
\begin{equation}
	y_{n}^{\rm F{(2)}}= \left( r_{{\rm R}n} + \mathbf{h}_{{\rm I}n}^H\mathbf{ \Theta }^{(2)}\mathbf{G}_{{\rm RI} }\right)x^{\rm F{(2)}}+ z_{n}^{(2)}.
\end{equation}
The received SINR to decode $s_{n}$ is given by
\begin{equation}
		\textup{SINR}_{n{\rightarrow} n}^{\rm F{(2)}}=\frac{\beta_{n} P_{r}| r_{{\rm R}n} + \mathbf{h}_{{\rm I}n}^H\mathbf{ \Theta }^{(2)}\mathbf{G}_{{\rm RI} }|^2}{\beta_{d} P_{r}| r_{{\rm R}n} + \mathbf{h}_{{\rm I}n}^H\mathbf{ \Theta }^{(2)}\mathbf{G}_{{\rm RI} }|^2+\sigma_{n}^2}.
\end{equation}
Similarly, due to the MRC, the received SINR at user $k\in\{n, d\}$ to decode the signal can be expressed as
	\begin{equation}
		\tilde{\gamma}^{\rm F}_{{k},\textup{MRC}}=\textup{SINR}_{k{\rightarrow}k}^{{(1)}}+\textup{SINR}_{k{\rightarrow}k}^{\rm F(2)}.
	\end{equation}
where $	\textit R^{\rm F}_{{k},\textup{MRC}}=\frac{1}{2}\log_2\left(1+\tilde{\gamma}^{\rm F}_{{k},\textup{MRC}}\right)$ is the data rate of user $k$ based on MRC principle. The final achievable data rate of user $k\in\{n, d\}$ can be written as
\begin{equation}
	\textit R^{\rm F}_{k}=\min\left\{\textit R_{r{\rightarrow}k}^{{(1)}}, \textit R^{\rm F}_{{k},\textup{MRC}}\right\}.
\end{equation}
As a result, the sum rate of the downlink transmission system employing F-NOMA scheme can be expressed as
\begin{equation}
	\textit R^{\rm F}_{{sum}}=\textit R^{\rm F}_{n}+\textit R^{\rm F}_{d}.
\end{equation}
\subsection{Problem Formulation }
To exploit the optimal performance for both communication protocols in the coexisting passive RIS and active relay assisted NOMA system, in this paper, we investigate the joint optimization of resource allocation based on proposed two protocols, including power allocation and RIS phase-shift design. Define $ {\boldsymbol \alpha}\triangleq\{\alpha_{n}, \alpha_{d}\}$, $ {\boldsymbol \beta}\triangleq\{\beta_{n}, \beta_{d}\}$ and ${\boldsymbol \Phi}\triangleq\{\mathbf{ \Theta }^{{(1)}},\mathbf{ \Theta }^{(2)}\}$, we consider the following problems.
\subsubsection{Problem for sum rate maximization}
The most common objective is to maximize the sum rate of all users. Given the target rates of paired users, the optimization problem of maximizing the system sum rate is formulated as
\begin{subequations}\label{P_max}
	\begin{align}
	&\mathop{\rm{max}}\limits_{ \boldsymbol{\alpha}, \boldsymbol {\beta},{\boldsymbol \Phi}}\quad
		\textit R_{sum}^{\rm{X}}\\
		&\label{alpha_max}\,\;\;{\rm s.t.} \;\;
		0\le \alpha_{n} \le \mathbf{\alpha}_{d}, \alpha_{n}+\alpha_{d}=1,\\
		&\label{beta_max}\,\quad\quad\;\,
	    0\le \beta_{d} \le \mathbf{\beta}_{n}, \beta_{n}+\beta_{d}=1,\\
		&\label{QoS_max}\,\quad\quad\;\,
		\textit R_{k}^{\rm{X}}\ge \textit R_{k}^{\rm \min}, \forall k\in\mathcal{K},\\
		&\label{Theta_max}\,\quad\quad\;\,
|\mathbf{ \Theta }^{{(1)}}_{mm}|=|\mathbf{ \Theta }^{(2)}_{mm}|=1, \forall m\in\mathcal{M},
	\end{align}
\end{subequations}
where $\rm{X}\in\{\rm{H, F}\}$ indicates the employed communication protocol. Constraints \eqref{alpha_max} and \eqref{beta_max} represent the power allocation requirements at the AP and DF relay, respectively. Constraint \eqref{beta_max} is only valid when the F-NOMA protocol is employed. Constraint \eqref{QoS_max} ensures that user $n$ and $d$ satisfy the target rate $\textit R_{n}^{\min}$ and $\textit R_{d}^{\min}$, respectively. Constraint \eqref{Theta_max} restricts unit-modulus phase shifters at the RIS.
\subsubsection{Problem for min rate maximization}
Our objective is to maximize the users’ communication performance in a fair manner. The max-min rate fairness problem is formulated as
	\begin{subequations}\label{P_min}
	\begin{align}
		&\mathop{\rm{max}}\limits_{ \boldsymbol{\alpha}, \boldsymbol {\beta},{\boldsymbol \Phi}}\;\,
		{\rm min} \{\textit R_{k}^{\rm X}, \forall k\in\mathcal{K}\}\\
		& \,\;\; {\rm s.t.} \;\;
		\eqref{alpha_max}, \eqref{beta_max}, \eqref{Theta_max}.
	\end{align}
\end{subequations}
Comparing the above two problems,  one key observation is that problem \eqref{P_min} contains fewer constraints. More precisely, problem \eqref{P_max} covers all the constraints of problem \eqref{P_min}. Therefore, this max-min problem can be solved in a similar manner as the problem for maximizing the system sum rate.
\subsection{Discussion}
Due to the more intricate constraints between the values of the optimization variables in two-stage transmission, which makes the resource allocation design of this considered system more complicated in practice than the conventional RIS-assisted networks. Focus on problem \eqref{P_max}, the formulated optimization problem for F-NOMA can be regarded as an upgrade of H-NOMA. In terms of solution difficulty, it is more challenging to coordinate the two-stage power allocation at the transmitter. Inspired by this, this paper is devoted to exploring the optimal resource allocation solutions based on F-NOMA protocol, while the similar method can be employed to address the problem for H-NOMA.

However, the original problem \eqref{P_max} is an intractable non-convex optimization problem that is difficult to solve directly by common standard optimization techniques. The main challenges can be summarized as follows:
1) problem \eqref{P_max} involves the coupling of multiple resource allocation coefficients (i.e. $\boldsymbol{\alpha}$, $\boldsymbol {\beta}$ and ${\boldsymbol \Phi}$); 2) the unit-modulus constraint \eqref{Theta_max} is non-convex, which further exacerbates the optimization difficulty.
In the following, we first develop an AO based iterative algorithm to find a high-quality suboptimal solution for the two considered  communication protocols in Section III. To further reduce computational complexity, a JO based approach is presented in Section IV. 
\section{AO Based Algorithm}
In this section, we firstly decompose problem \eqref{P_max} for F-NOMA transmission into two subproblems, i.e., power allocation optimization and RIS phase-shift optimization. Then, an AO based algorithm is applied to solve these subproblems one by one until convergence.
\subsection{Power Allocation Optimization}	
First, we optimize the power allocation coefficients $\{ \boldsymbol{\alpha},  \boldsymbol{\beta}\}$ with fixed RIS phase-shift matrices $\{\mathbf \Theta^{(1)},\mathbf \Theta^{(2)}\}$. Problem \eqref{P_max} is simplified as
\begin{subequations}\label{P_PA}
	\begin{align}
		&\mathop{\rm{max}}\limits_{ \boldsymbol{\alpha},  \boldsymbol{\beta}}\quad
		\textit R_{sum}^{\rm{F}}\\
		&\label{SINR_PA}\;\;{\rm s.t.} \;\;
		\textup{SINR}_{k}^{\rm{F}}\ge {\boldsymbol \gamma}_{\textit k}^{\rm \min},  \forall k\in\mathcal{K},\\
		&\quad\quad\;\,
		\eqref{alpha_max}, \eqref{beta_max},
	\end{align}
\end{subequations}
where constraint \eqref{SINR_PA} is a variant of constraint \eqref{QoS_max}. $\textup{SINR}_{k}^{\rm F}\triangleq\min\big\{{\textup{SINR}}_{r{\rightarrow}k}^{{(1)}},\big({\textup {SINR}}_{k{\rightarrow} k}^{(1)}+{\textup {SINR}}^{{\rm F}(2)}_{k{\rightarrow} k}\big)\big\}$ denotes the effective SINR received at user $k\in\{n, d\}$. $\boldsymbol{\gamma}_ k^{\rm min}= 2^{2\textit R_ k^{\rm min}}-1$ denotes the predefined target SINR of signal $s_k$ required to support successful SIC execution at user $k$.

To derive feasible solutions for problem \eqref{P_PA}, which is equal to find feasible sets of optimization variables $\alpha_{n}$ and $\beta_{d}$, since $\alpha_{d}=1-\alpha_{n}$ and $\beta_{n}=1-\beta_{d}$. For ease of exposition, we define
\begin{equation}
	\begin{aligned}		
		&\Gamma_{{\rm AR}}\triangleq\frac{P_{a}|r_{{\rm AR} } + \mathbf{h}_{\rm{IR}}^H\mathbf{ \Theta }^{(1)}\mathbf{G}_{{\rm AI} }|^2}{\sigma_r^2},\\
		&\Gamma_{{\rm A}k}\triangleq\frac{P_{a}|r_{{\rm A}k} +  \mathbf{h}_{{\rm I}k}^H\mathbf{ \Theta }^{(1)} \mathbf{G}_{{\rm AI} }|^2}{\sigma_k^2},\\
		&\Gamma_{{\rm R}k}\triangleq\frac{P_{r}|r_{{\rm R}k} + \mathbf{h}_{{\rm I}k}^H\mathbf{ \Theta }^{(2)}\mathbf{G}_{{\rm RI} }|^2}{\sigma_k^2},
	\end{aligned}
\end{equation}
where $r_{{\rm A}d}=0$ since there is no direct link between the AP and user $d$.

Then, we need to clarify the feasibility conditions of the subproblem to ensure that there is at least one feasible solution to the problem. It can be concluded that $\alpha_n$ should satisfy $\alpha_n^{\min} \le\alpha_n\le\alpha_n^{\max}$, meanwhile, $\beta_d$ should satisfy $\beta_d^{\min} \le\beta_d\le\beta_d^{\max}$. To elaborate  the feasibility conditions of problem \eqref{P_PA}, the following theorem is presented.
\begin{theorem}
The subproblem is feasible if and only if the following conditions hold.
\begin{align}
	\textup{Condition 1:} \;\alpha_n^{\min}\le\alpha_n^{\max},\\
	\textup{Condition 2:} \;\beta_d^{\min}\le\beta_d^{\max},
	\end{align}
where $\alpha_n^{\min}$, $\alpha_n^{\max}$, $\beta_d^{\min}$ and $\beta_d^{\max}$ are expressed, respectively, as
\begin{align}
 &\left\{
 \begin{aligned}
 &\alpha_n^{\min}= \frac{\boldsymbol{\gamma}_n^{\rm min}}{\Gamma_{{\rm AR}}}, \\ 
 &\alpha_n^{\max}=\min\big\{\frac{1}{2},\frac{{\Gamma_{{\rm AR}}}-{\boldsymbol{\gamma}_d^{\rm min}}}{{\Gamma_{{\rm AR}}}({\boldsymbol{\gamma}_d^{\rm min}}+1)}\big\},\end{aligned}
 \right. \textup{and}\\
 &\left\{
 \begin{aligned}
 &\beta_d^{\min}= \max\big\{0,\frac{\boldsymbol{\gamma}_d^{\rm min}}{\Gamma_{{\rm A}d}}-\frac{(1-\alpha_n)\Gamma_{{\rm A}d}}{\Gamma_{{\rm R}d}(\alpha_n\Gamma_{{\rm A}d}+1)}\big\},\\ &\beta_d^{\max}=\min\big\{\frac{1}{2},\frac{{\Gamma_{{\rm R}n}}+1}{(\boldsymbol{\gamma}_n^{\rm min}-\alpha_n\Gamma_{{\rm A}n}+1)\Gamma_{{\rm R}n}}-\frac{1}{\Gamma_{{\rm R}n}}\big\}.\end{aligned}
\right. 
\end{align}
\begin{proof}
	From the target SINR constraint at user $n$ as $\min\big\{{\textup {SINR}}_{r{\rightarrow}n}^{{(1)}},\big({\textup{SINR}}_{n{\rightarrow} n}^{{(1)}}+{\textup {SINR}}^{{\rm F}{(2)}}_{n{\rightarrow}n}\big)\big\}\ge {\boldsymbol \gamma}_{n}^{\rm \min}$, we have 
\begin{align}
		&\mathbf{\alpha}_n\ge\frac{\boldsymbol \gamma_n^{\rm min}}{\Gamma_{{\rm AR}}},\\
		&\alpha_n\ge\frac{\boldsymbol \gamma_n^{\rm min}}{\Gamma_{{\rm A}n}}-\frac{\beta_n\Gamma_{{\rm R}n}}{\Gamma_{{\rm A}n}(\beta_d\Gamma_{{\rm R}n}+1)}.
\end{align}	
	Substituting $\beta_{n}=1-\beta_{d}$ into (34), which is equivalently rewritten as
$\beta_d\le\frac{{\Gamma_{{\rm R}n}}+1}{(\boldsymbol{\gamma}_n^{\rm min}-\alpha_n\Gamma_{{\rm A}n}+1)\Gamma_{{\rm R}n}}-\frac{1}{\Gamma_{{\rm R}n}}$.
As for the target SINR constraint at user $d$, $\min\big\{{\textup {SINR}}_{r{\rightarrow}d}^{{(1)}},\big({\textup {SINR}}_{d{\rightarrow}d}^{(1)}+{\textup {SINR}}^{{\rm F}{(2)}}_{d{\rightarrow} d}\big)\big\}\ge {\boldsymbol \gamma}_{d}^{\min}$, implying 
\begin{align}
    &\alpha_d\ge{\boldsymbol\gamma_d^{\min}} \alpha_n+\frac{\boldsymbol\gamma_d^{\rm min}}{\Gamma_{{\rm AR}}},\\
    &\beta_d\ge\frac{\boldsymbol\gamma_d^{\rm min}}{\Gamma_{{\rm R}d}}-\frac{\alpha_d\Gamma_{{\rm A}d}}{\Gamma_{{\rm R}d}(\alpha_n\Gamma_{{\rm A}d}+1)}.
\end{align}	
Substituting $\alpha_{d}=1-\alpha_{n}$ into (35) and (36), we obtain
$\alpha_n\le\frac{{\Gamma_{{\rm AR}}}-{\boldsymbol{\gamma}_d^{\rm min}}}{{\Gamma_{{\rm AR}}}({\boldsymbol{\gamma}_d^{\rm min}}+1)}$ and $\beta_{d}\ge\frac{\boldsymbol{\gamma}_d^{\rm min}}{\Gamma_{{\rm A}d}}-\frac{(1-\alpha_n)\Gamma_{{\rm A}d}}{\Gamma_{{\rm R}d}(\alpha_n\Gamma_{{\rm A}d}+1)}$, respectively.
With the results above, all SINR boundaries in the theorem can be transformed into expressions only for $\alpha_{n}$ and $\beta_{d}$. In addition, constraints \eqref{alpha_max} and \eqref{beta_max} notify that, $0\le\alpha_n\le1/2$ and $0\le\beta_d\le1/2$, which completes the proof.
\end{proof}

\end{theorem}
Due to (32), the value of $\alpha_n$ determines the feasible solution of $\beta_d$.
Assuming that problem \eqref{P_PA} is feasible, i.e., conditions (29) and (30) hold, let $\mathcal{S}$ and $\mathcal{G}$ denote the feasible set of $\alpha_n$ and $\beta_d$ for problem, respectively, which are given by
\begin{align}
&	\mathcal{S}=\{\alpha_n|\alpha_n^{\min} \le\alpha_n\le\alpha_n^{\max}\},\\
&	\mathcal{H}=\{\beta_d|\beta_d^{\min} \le\beta_d\le\beta_d^{\max},\forall\alpha_n\in\mathcal{S}\}.
\end{align}
The feasible schemes of power allocation optimization are presented as  
\begin{equation}
	\mathcal{P}=\{(\alpha_n,\beta_d)|\forall\alpha_n\in\mathcal{S},\forall\beta_d\in\mathcal{H}\}.
	\end{equation}
Among them, the scheme that achieves the maximum objective value is selected as the optimal scheme for power allocation as follows:
\begin{equation}
	\mathbf{p}^{\rm F*}=(\alpha_n^*,\beta_d^{*})=\arg\max_{\mathbf{p}^{\rm F}\in\mathcal{P}} \textit R_{sum}^{\rm{F}}(\mathbf{p}^{\rm F}).
\end{equation}
Note that,  however, we cannot directly judge the monotonicity of $\textit R_{sum}^{\rm{F}}(\alpha_n,\beta_d)$ with respect to either $\alpha_n$ or $\beta_d$. To obtain the optimal solution $\mathbf{p}^{\rm F*}$, a straightforward approach is to exhaustively two-dimensional search over all possible power allocation schemes and select the best candidate solution. 
\begin{remark} 	
    For H-NOMA, the power allocation subproblem is reduced to a one-dimensional search for all possible $\alpha_\textit n$ to find the optimal solution.		
\end{remark} 
\subsection{RIS Phase-Shift Optimization}
With given power allocation coefficients $\{ \boldsymbol{\alpha},  \boldsymbol{\beta}\}$, the subproblem for RIS phase-shift optimization is reduced into
\begin{subequations}\label{P_RIS}
	\begin{align}
		&\mathop{\rm{max}}\limits_{\boldsymbol \Phi}\quad
			 \textit R_{sum}^{\rm{F}}\\
		&\;\;{\rm s.t.} \;\;
		|\mathbf{ \Theta }^{{(1)}}_{mm}|=|\mathbf{ \Theta }^{(2)}_{mm}|=1, \forall m\in\mathcal{M},\\
		&\quad\quad\;\,
    	\eqref{SINR_PA}.
	\end{align}
\end{subequations}
More explicitly, we consider jointly optimizing the passive beamforming of two stages.
To transform problem \eqref{P_RIS} into a more tractable form, define  
\begin{align}
	\mathbf Z_{i}\triangleq
	\begin{bmatrix}
		\mathbf Q_{i}     \\
		r_{i}^H  
	\end{bmatrix}
	, i\in \{{\rm AR}, \textup{A}k, {\rm R}k\}, k\in\{n, d\}
\end{align}
where $\mathbf Q_{\textup {AR}}={\textup{diag}}(\mathbf{h}_{\rm{IR}}^H)\mathbf{G}_{{\rm AI} }\in\mathbb{C}^{M \times 1}$, 
$\mathbf Q_{{\rm A}k}={\textup {diag}}(\mathbf{h}_{{\rm I}k}^H)\mathbf{G}_{{\rm AI} }\in\mathbb{C}^{M \times 1}$,
and $\mathbf Q_{{\rm R}k}={\textup{diag}}(\mathbf{h}_{{\rm I}k}^H)\mathbf{G}_{{\rm RI} }\in\mathbb{C}^{M \times 1}$ are the cascade channels assisted by RISs.
Let $\mathbf{v}_1=[v_1^{(1)},v_2^{(1)},...,v_M^{(1)}]$, $\mathbf{v}_2=[v_1^\textit{(2)},v_2^\textit{(2)},...,v_M^\textit{(2)}]$, $ \mathbf {w}_1=[\mathbf v_1,1]$,  $ \mathbf {w}_2=[\mathbf v_2,1]$, where $v_m^{(1)}=e^{j\theta_{m}^{(1)}}$, $v_m^{(2)}=e^{j\theta^{(2)}_m}, \forall m\in\mathcal{M}$. Based on above definitions, we transform problem \eqref{P_RIS} into
 \begin{subequations}\label{P_RIS2}
 	\begin{align}
 		&\mathop{\rm{max}}\limits_{\boldsymbol \Phi}\quad
 	  	\textit R_{sum}^{\rm{F}}\\
 		&\label{SINR_N}\;\;{\rm s.t.}\, 
 			\min\Big\{ \frac{\alpha_{n} P_{a}|\mathbf {w}_1 \mathbf Z_{{\rm AR}}|^2}{\sigma_r^2},  \big( \frac{\alpha_{n} P_{a}|\mathbf {w}_1 \mathbf Z_{{\rm A}n}|^2}{\sigma_n^2}\notag\\
 		&\quad\;\;\;+\frac{\beta_{n} P_{r}|\mathbf {w}_2 \mathbf Z_{{\rm R}n}|^2}{\beta_{d}P_{r}|\mathbf {w}_2 \mathbf Z_{{\rm R}n}|^2+\sigma_n^2}\big) \Big\}\ge \boldsymbol{\gamma}_n^{\rm min},\\ 
 		&\label{SINR_D}\;\;\quad\;\,\min\Big\{\frac{\alpha_{d}P_{a}|\mathbf {w}_1 \mathbf Z_{{\rm AR}}|^2}{\alpha_{n}P_{a}|\mathbf {w}_1 \mathbf Z_{{\rm AR}}|^2+\sigma_r^2}, \big(\frac{\alpha_{d}P_{a}|\mathbf {w}_1 \mathbf Z_{{\rm A}d}|^2}{\alpha_{n}P_{a}|\mathbf {w}_1 \mathbf Z_{{\rm A}d}|^2+\sigma_d^2}\notag\\
 		&\quad\;\;\;+\frac{\beta_{d}P_{r}|\mathbf {w}_2 \mathbf Z_{{\rm R}d}|^2}{\sigma_d^2}\big)\Big\}\ge \boldsymbol{\gamma}_d^{\rm min},\\ 	
 		&\label{SHIFT_V}\;\;\quad\;\,
 		|v_m^{(1)}|=|v_m^{(2)}|=1, \forall m\in\mathcal{M}.	
 	\end{align}
 \end{subequations}
 
To tackle the unit-modulus constraint \eqref{SHIFT_V}, we lift $\{\mathbf{w}_t, t \in\{1,2\}\}$ into two positive semidefinite (PSD) matrices $\mathbf {W}_t\in\mathbb{C}^{(M+1)\times (M+1)}$, satisfying $\mathbf W_t=\mathbf {w}^H_t\mathbf{w}_t$ and  $\textup{Rank} (\mathbf W_t)=1$, then problem \eqref{P_RIS2} can be rewritten as problem \eqref{P_RIS3}, which are shown in the top of the next page.
\begin{figure*}[!t]
	\normalsize
	\begin{subequations}\label{P_RIS3}
	\begin{align}
		&\mathop{\rm{max}}\limits_{\mathbf{W}_t}\quad
			\textit R_{sum}^{\rm{F}}\\
		&\label{SINR_N2}\;\;{\rm s.t.}\, 
		\min\Big\{ \frac{\alpha_{n} P_{a}\textup{Tr}(\mathbf {W}_1 \mathbf Z_{ {\rm AR}}\mathbf Z^H_{{\rm AR}})}{\sigma_r^2}, \frac{\alpha_{n} P_{a}\textup{Tr}(\mathbf {W}_1 \mathbf Z_{{\rm A}n}\mathbf Z^H_{{\rm A}n})}{\sigma_n^2}\!+\!\frac{\beta_{n} P_{r}\textup{Tr}(\mathbf {W}_2 \mathbf Z_{{\rm R}n}\mathbf Z^H_{{\rm R}n})}{\beta_{d}P_{r}\textup{Tr}(\mathbf {W}_2 \mathbf Z_{{\rm R}n}\mathbf Z^H_{{\rm R}n})\!+\!\sigma_n^2} \Big\}\ge \boldsymbol{\gamma}_n^{\rm min},\\	
		&\label{SINR_D2}\;\;\quad\;\,  
		\min\Big\{ \frac{\alpha_{d} P_{a}\textup{Tr}(\mathbf {W}_1 \mathbf Z_{{\rm AR}}\mathbf Z^H_{{\rm AR}})}{\alpha_{n}P_{a}\mathrm{Tr}(\mathbf {W}_1 \mathbf Z_{{\rm AR}}\mathbf Z^H_{{\rm AR}})+\sigma_r^2},\frac{\alpha_{d} P_{a}\textup{Tr}(\mathbf {W}_1 \mathbf Z_{{\rm A}d}\mathbf Z^H_{{\rm A}d})}{\alpha_{n}P_{a}\textup{Tr}(\mathbf {W}_1 \mathbf Z_{{\rm A}d}\mathbf Z^H_{{\rm A}d})+\sigma_d^2}+
		\frac{\beta_{d}P_{r}\textup{Tr}(\mathbf {W}_2 \mathbf Z_{{\rm R}d}\mathbf Z^H_{{\rm R}d})}{\sigma_d^2}\Big\}\ge \boldsymbol{\gamma}_d^{\rm min},\\ 	
		&\label{Succeq}\;\;\quad\;\,\mathbf W_t\succeq 0, t\in\{1,2\},\\
		&\label{Rank}\;\;\quad\;\,	\textup{Rank}(\mathbf {W}_t)=1, t\in\{1,2\},\\
		&\label{SHIFT_W}\;\;\quad\;\,
		[\mathbf {W}_t]_{mm}=1, \forall m \in \{1,...,M+1\}, t\in\{1,2\}.
		\end{align}
\end{subequations}
\hrulefill \vspace*{0pt}
\end{figure*}
As a result, this problem remains non-convex and challenging to deal with due to the non-convexity in \eqref{SINR_N2}, \eqref{SINR_D2} and \eqref{Rank}. 
To guarantee rank-one property of $\mathbf {W}_t$, we adopt the difference-of-convex (DC) relaxation method to extract the rank-one solution from high-rank matrix as~\cite{STAR_Xidong}. To elaborate it, an equivalent representation of $\textup{Rank}(\mathbf {W}_t)=1$ is provided as
\begin{align}
	\|\mathbf  {W}_t\|_*-	\|\mathbf {W}_t\|_2=0, t\in\{1,2\},
\end{align}
where $\|\mathbf {W}_t\|_*=\sum_j\sigma_j(\mathbf {W}_t)$ and $\|\mathbf  {W}_t\|_2= \sigma_1(\mathbf  {W}_t)$ denote the nuclear norm and spectral
norm, respectively, and $\sigma_j(\mathbf {W}_t)$ is the $j$-th largest singular value of matrix $\mathbf {W}_t$.

Then, a penalty based DC programming is employed to optimize the RIS phase-shift. Inspired by (45), if we make the value of $\|\mathbf {W}_t\|_*-	\|\mathbf {W}_t\|_2$ as small as possible, the rank of the matrix $\{\mathbf {W}_t, t\in\{1,2\}\}$ can be approximated to one. By substituting the above representation into the objective function of problem \eqref{P_RIS3}, we have
\begin{subequations}\label{P_RIS4}
	\begin{align}
		& \mathop{\rm{min}}\limits_{{\mathbf {W}_t}}	\quad
		-\textit R_{sum}^{\rm{F}}+\frac{1}{\eta}  \sum_{t\in\{1,2\}}(\|\mathbf {W}_ t\|_*-	\|\mathbf {W}_t\|_2)\\
		&\;\;{\rm s.t.} \;\;
		\eqref{SINR_N2},\eqref{SINR_D2},\eqref{Succeq},\eqref{SHIFT_W}.
	\end{align}
\end{subequations}
where equality constraint (45) is relaxed to a penalty term, and $\eta>0$ is the penalty factor which penalizes the objective function if $\{\mathbf{W}_t\}$ is not rank-one. By enforcing the penalty term to be zero, problem \eqref{P_RIS4} induces two exact rank-one matrices. 
It can be verified that, when  $\eta\to 0$, the optimal solution $\{\mathbf {W}_t^{\ast}\}$ of the problem always satisfies the equality constraint (45), i.e., problems \eqref{P_RIS3} and \eqref{P_RIS4} are equivalent~\cite{Penalty}. 

To proceed, we resort to the SCA technique~\cite{Crashworthiness} to construct a convex upper approximation function of the penalty term as follows
\begin{equation} 
	\begin{split}
		\|\mathbf {W}_t\|_*-	\|\mathbf {W}_t\|_2 \le\|\mathbf {W}_ t\|_*-\overline{\mathbf {W}}^{l}_t, 
	\end{split}
\end{equation} 
where $\overline{\mathbf {W}}^{l}_t\triangleq \|\mathbf {W}^{l}_ t\|_2+\textup{Tr}\big(\xi_{\rm max}(\mathbf {W}^{l}_\textit t)\xi_{\rm max}(\mathbf {W}^{l}_ t)^H(\mathbf {W}_ t-\mathbf {W}^{l}_ t)\big)$. $\mathbf {W}_t^{l}$ is a given point  in the $l$-th iteration, $\xi_{\rm max}(\mathbf{W}^{l}_t)$ denotes the eigenvector corresponding to the largest eigenvalue of $\mathbf {W}^{l}_t$. Substituting (47) into the objective function of problem \eqref{P_RIS4}, we can further obtain the objective function as
	\begin{equation}
		\mathop{\rm{min}}\limits_{{\mathbf {W}_t}}\quad 
		-\textit R_{sum}^{\rm{F}}+\frac{1}{\eta}  \sum_{t\in\{1,2\}}(\|\mathbf {W}_t\|_*-\overline{\mathbf {W}}^{l}_t).
	\end{equation}
 Up to this point, problem \eqref{P_RIS4} is still intractable due to the non-convex target SINR constraints \eqref{SINR_N2} and \eqref{SINR_D2}. To overcome this issue, we introduce the slack variables $X_{{\rm R}n}$ and $Y_{{\rm R}n}$ to rewrite $\textup{SINR}_{n{\rightarrow}n}^{{\rm F}{(2)}}$ into a binary function as in {\textup{\textbf{Lemma 1}}}, i.e.,
\begin{equation}
	\textup{SINR}_{n{\rightarrow}n}^{{\rm F}{(2)}}= \frac{1}{X_{{\rm R}n}Y_{{\rm R}n}},
\end{equation}
where $
	\frac{1}{X_{{\rm R}n}}={\beta_{n}P_{r}\textup{Tr}(\mathbf {W}_2 \mathbf Z_{{\rm R}n}\mathbf Z^H_{{\rm R}n})}$ and $Y_{{\rm R}n}=
	{\beta_{d}P_{r}\textup{Tr}(\mathbf {W}_2 \mathbf Z_{{\rm R}n}\mathbf Z^H_{{\rm R}n})+\sigma_n^2}$.
\begin{lemma} For $X \ge 0$ and $Y \ge 0$, $f(X,Y)=\frac{1}{XY}$ is a joint convex function with respect to $X$ and $Y$. 
	\begin{proof} It is easy to prove {\textup{\textbf{Lemma 1}}}  by showing that the Hessian matrix of function $f(X,Y)$ is positive semidefinite when $X,Y \ge 0$. Therefore, $f(X,Y)$ is a convex function.
	\end{proof}
\end{lemma}

Recall that any convex function is globally lower bounded by its first-order Taylor expansion at any point~\cite{convex}. Then, the SCA technique can be leveraged to obtain the lower bound of $\textup{SINR}_{n{\rightarrow}n}^{{\rm F}{(2)}}$, which is approximated by a more tractable function with given local points $\{ X^{l}_{{\rm R}n},Y^{l}_{{\rm R}n}\}$ in the $l$-th iteration, as
\begin{equation}
	\begin{aligned} 
	\textup{SINR}_{n{\rightarrow}n}^{{\rm F}{(2)}}&\ge \frac{1}{X^{l}_{{\rm R}n}Y^{l}_{{\rm R}n}}-
	\frac{X_{{\rm R}n}-X^{l}_{{\rm R}n}}{{X^{l}_{{\rm R}n} }^2Y^{l}_{{\rm R}n}}
	-\frac{Y_{{\rm R}n}-Y^{l}_{{\rm R}n}}{{Y^{l}_{{\rm R}n} }^2X^{l}_{{\rm R}n}}\\
	&\triangleq [{\mathfrak{T}}_{n{\rightarrow}n}^{{\rm F}{(2)}}]^{\textup{lower}}.
		\end{aligned}
\end{equation}
In a similar way, introduce $X_{{\rm AR}}=\frac{1}{\alpha_{d}P_{a}\textup{Tr}(\mathbf {W}_1\mathbf Z_{{\rm AR} } \mathbf Z^H_{{\rm AR}})}$, $Y_{{\rm AR}}={\alpha_{n}P_{a}\textup{Tr}(\mathbf {W}_1 \mathbf Z_{{\rm AR}}\mathbf Z^H_{{\rm AR}})+\sigma_r^2}$ and $X_{{\rm A}d}=\frac{1}{\alpha_{d}P_{a}\textup{Tr}(\mathbf {W}_1\mathbf Z_{{\rm A}d} \mathbf Z^H_{{\rm A}d})}$ and $Y_{{\rm A}d}={\alpha_{n}P_{a}\textup{Tr}(\mathbf {W}_1 \mathbf Z_{{\rm A}d}\mathbf Z^H_{{\rm A}d})+\sigma_d^2}$ to replace $X_{{\rm R}n}$, $Y_{{\rm R}n}$, respectively, and then perform a first-order Taylor expansion like (50), the concave lower bound expressions of $\textup{SINR}_{r{\rightarrow}d}^{{(1)}}$ and $\textup{SINR}_{d{\rightarrow}d}^{(1)}$ can be obtained, respectively, denoted by $[{\mathfrak{T}}_{r{\rightarrow}d}^{{(1))}}]^{\textup{lower}} $ and $[{\mathfrak{T}}_{d{\rightarrow}d}^{(1)}]^{\textup{lower}}$.

With the results above, the subproblem for RIS phase-shift optimization can be transformed into as
 \begin{subequations}\label{P_RIS5}
	\begin{align}
		&\mathop{\rm{min}}\limits_{{\mathbf {W}_t}}\quad 
		-\textit R_{sum}^{\rm{F}}+\frac{1}{\eta}  \sum_{t\in\{1,2\}}(\|\mathbf {W}_t\|_*-\overline{\mathbf {W}}^{l}_t)\\
		&\;\;{\rm s.t.} 
		\min\Big\{ \frac{\alpha_{n} P_{a}\textup{Tr}(\mathbf {W}_1 \mathbf Z_{\textup {AR}}\mathbf Z^H_{\textup {AR}})}{\sigma_r^2},\notag\\
		&\qquad\frac{\alpha_{n} P_{a}\textup{Tr}(\mathbf {W}_1 \mathbf Z_{{\rm A}n}\mathbf Z^H_{{\rm A}n})}{\sigma_n^2}+[{\mathfrak{T}}_{n{\rightarrow}n}^{{\rm F}{(2)}}]^{\textup{lower}}\Big\} \ge \boldsymbol{\gamma}_n^{\rm min},\\	
		&\qquad \min\Big\{[{\mathfrak{T}}_{r{\rightarrow}d}^{(1)}]^{\textup{lower}},\notag\\
		&\qquad [{\mathfrak{T}}_{d{\rightarrow}d}^{(1)}]^{\textup{lower}}+
		\frac{\beta_{d}P_{r}\textup{Tr}(\mathbf {W}_2 \mathbf Z_{{\rm R}d}\mathbf Z^H_{{\rm R}d})}{\sigma_d^2}\Big\} \ge\boldsymbol{\gamma}_d^{\rm min},\\ 	
    	&\qquad
		\eqref{Succeq},\eqref{SHIFT_W}.
	\end{align}
\end{subequations}
For any given $\{{\boldsymbol\alpha}, {\boldsymbol\beta}\}$, the rank-relaxed problem \eqref{P_RIS5} is jointly convex with respect to $\mathbf {W}_1$ and $\mathbf {W}_2$, which can be optimally solved by existing convex optimization solvers such as CVX~\cite{cvx}.
It is noteworthy that our proposed penalty based DC algorithm comprises two layer iterations:
\begin{itemize}
	\item 
Inner layer iteration: With the given penalty factor, $\{\mathbf W_t, t\in\{1,2\}\}$ are jointly optimized by iteratively solving the relaxed problem \eqref{P_RIS5}. 
	\item 
Outer layer iteration: We gradually decrease the value of the penalty factor $\eta$ as follows:
\begin{equation}
	\eta=c\eta,
\end{equation}
where $c $ $(0 < c < 1)$ is a scaling factor, where a larger value of $c$ can achieve better performance but at the cost of more iterations in the outer layer. Eventually, the algorithm terminates when the penalty term satisfies the following criterion:
\begin{equation}
	\max\big\{\|\mathbf {W}_t\|_*-	\|\mathbf {W}_t\|_2, t\in\{1, 2\}\big\}\le \epsilon,
\end{equation}
where $\epsilon$ denotes a predefined maximum violation of equality constraint (45). 
\end{itemize}
The details of the developed algorithm are summarized in \textbf{Algorithm 1}.
\begin{algorithm}[!t]\label{method1}
	\caption{Proposed Penalty Based DC Algorithm for Solving Problem \eqref{P_RIS}}
	\begin{algorithmic}[1]
		\STATE {{Initialize} $\big\{\mathbf{W}_t^{(0)}, t\in\{1, 2\}\big\}$, $\eta$, $r_1=0$, $r_2=0$,  $\epsilon $, $\varepsilon$, $r_{max}$.}
		\STATE {\bf repeat: outer loop}
		\STATE \quad Set iteration index $r_1\leftarrow0$ for inner loop.
		\STATE \quad {\bf repeat: inner loop}
		\STATE \quad\quad Solve problem \eqref{P_RIS5} with given power allocation solution, and update $\{\mathbf{W}_t^{(*)}\}=\{\mathbf{W}_t^{( r_1+1)}\}$.
		\STATE \quad\quad $r_1\leftarrow r_1+1$.
		\STATE \quad {\bf until} the fractional decrease of the objective function value is below a predefined threshold $ \varepsilon >0$ or the maximum number of inner iterations $r_{\max}$ is reached.
		\STATE \quad Update $\{ \mathbf{W}_t^{(0)}\}$ with the current solutions $\{\mathbf{W}_t^{(r_1)}\}$.
		\STATE \quad Update penalty factor $\eta^{(r_2)}\leftarrow c\eta^{(r_2)}$ based on (52).
		\STATE \quad $r_2\leftarrow r_2+1$.
		\STATE {\bf until} the constraint violation is below a predefined threshold $\epsilon>0$.
	\end{algorithmic}
\end{algorithm}
\begin{remark} 
	For H-NOMA, the closed-form IR-stage RIS phase-shift solution can be derived based on the following triangle inequality,
	\begin{equation}
		|r_{{\rm R}d} + \mathbf{h}_{{\rm I}d}^H \mathbf{ \Theta }^{(2)}\mathbf{G}_{{\rm RI}}|\le|r_{{\rm R}d}|+ |\mathbf{h}_{{\rm I}d}^H \mathbf{ \Theta}^{(2)}\mathbf{G}_{{\rm RI}}|,
	\end{equation}
	 we can always obtain a reflection coefficients matrix $\mathbf{ \Theta}^{(2)}$ that satisfies (54) with equality. The optimal solution to $m$-th phase-shifter at the RIS can be expressed as
	\begin{equation}
		\begin{aligned}
			\theta_{m}^ {(2)*}&=\omega_0-\arg(\mathbf{h}_{{{\rm I}d},m}^H\mathbf{G}_{{{\rm RI}},m})\\&=\omega_0-\arg(\mathbf{h}_{{{\rm I}d},m}^H)-\arg(\mathbf{G}_{{{\rm RI}},m}),
		\end{aligned}
	\end{equation}
	where $\arg(\cdot)$ is the phase operator, $\mathbf{h}_{{{\rm I}d},m}^H$ denotes the $m$-th element of $\mathbf{h}_{{{\rm I}d}}^H$, and $\mathbf{G}_{{{\rm RI}},m}$ denotes the $m$-th element of $\mathbf{G}_{{{\rm RI}}}$, and the optimal phase-shift matrix $\mathbf{ \Theta}^{(2)*}$ is formed from $\theta_{m}^{(2)*}$. 
	(55) suggests that the configuration scheme for RIS is to achieve phase alignment of the cascaded channel and direct channel from the DF relay to user $d$.
\end{remark} 
\subsection{Proposed Algorithm, Convergence, and Complexity}
Based on the above two subproblems, we design an AO based algorithm for solving problrm \eqref{P_max} for F-NOMA. By solving subproblem \eqref{P_PA} and subproblem \eqref{P_RIS5}, the power allocation coefficients and passive RIS phase-shift are optimized alternately, and the solution obtained after each iteration are used as the input local point of the next iteration. To facilitate the understanding of the proposed algorithm, we summarize the details in \textbf{Algorithm 2}. It is noted that, since the proposed algorithm is an iterative method, a strictly feasible starting point is a prerequisite, especially during the iterative process in steps 3 and 4.
\begin{algorithm}[!t]\label{method1}
	\caption{Proposed AO based Algorithm for Solving Problem \eqref{P_max} for F-NOMA}
	\label{alg:A}
	\begin{algorithmic}[1]
		\STATE {Initialize feasible points $\big\{\mathbf{W}_t^{(0)}, t\in\{1, 2\}\big\}$ and  $\{{\boldsymbol\alpha^{(0)}}, {\boldsymbol\beta^{(0)}}\}$, $n=0$, $ \varepsilon$.}
		\REPEAT 
		\STATE {Exhaustively search for the optimal solution $\mathbf{p}^{\rm F*}$ according to (40) with given $\{\mathbf{W}_t^{(n)}\}$, and update $\{{\boldsymbol\alpha^{(*)}}, {\boldsymbol\beta^{(*)}}\}=\{{\boldsymbol\alpha^{(n+1)}}, {\boldsymbol\beta^{(n+1)}}\}$}.
		\STATE {Perform \textbf{Algorithm 1} to solve problem \eqref{P_RIS5} with given $\{{\boldsymbol\alpha^{(n+1)}}, {\boldsymbol\beta^{(n+1)}}\}$, and update $\{\mathbf{W}_t^{(*)}\}=\{\mathbf{W}_t^{(n+1)}\}$}.\\
		\STATE Update $n \leftarrow n+1$.\\
		\UNTIL the fractional increase of the objective value is below a threshold $\varepsilon$.\\
		\STATE	{Output} the optimal solutions $\{{\boldsymbol\alpha^{(*)}}, {\boldsymbol\beta^{(*)}}\}$ and $\{\mathbf{W}_t^{(*)}\}$.
	\end{algorithmic}
\end{algorithm}
\subsubsection{Convergence analysis}
The proposed \textbf{Algorithm 2} is guaranteed to converge  over the non-decreasing iterations~\cite{SCA} as the analysis shown in the following.
To prove the achievable sum rate obtained in \textbf{Algorithm 2} is monotonically non-decreasing, let us denote $\mathcal{R}\big(\{{\boldsymbol\alpha^{(n)}}, {\boldsymbol\beta^{(n)}}\}, \{\mathbf{W}_t^{(n)}\}\big)$ as the objective function’s value of problem \eqref{P_max} in the $n$-th iteration. Firstly, according to step 3 about the optimization of $\{{\boldsymbol\alpha^{(n)}}, {\boldsymbol\beta^{(n)}}\}$ for any given $\{{\mathbf{W}}_t^{(n)}\}$, we have:
        \begin{equation}	
		\mathcal{R}\big(\{{\boldsymbol\alpha^{(n)}}, {\boldsymbol\beta^{(n)}}\}, \{\mathbf{W}_t^{(n)}\}\big)
		\overset{(a)} \le
		\mathcal{R}\big(\{{\boldsymbol\alpha^{(n+1)}}, {\boldsymbol\beta^{(n+1)}}\}, \{\mathbf{W}_t^{(n)}\}\big),
		\end{equation}
	where $(a)$ holds since for given $\{{\mathbf{W}}_t^{(n)}\}$,
	$\{{\boldsymbol\alpha^{(n+1)}}, {\boldsymbol\beta^{(n+1)}}\}$ is the optimal solution to problem \eqref{P_PA} among all candidate solutions. 
   Based on step 4, the resulting mathematical expressions are as follows:
	\begin{equation}
	\begin{aligned}	
   			&\mathcal{R}\big(\{{\boldsymbol\alpha^{(n+1)}}, {\boldsymbol\beta^{(n+1)}}\}, \{\mathbf{W}_t^{(n)}\}\big)\\ &\overset{(b)}=\mathcal{R}^{\textup{lower}}_{\{\mathbf{W}_t^{(n)}\}}\big(\{{\boldsymbol\alpha^{(n+1)}}, {\boldsymbol\beta^{(n+1)}}\}, \{\mathbf{W}_t^{(n)}\}\big)\\
   			&\overset{(c)}\le \mathcal{R}^{\textup{lower}}_{\{\mathbf{W}_t^{(n)}\}}\big(\{{\boldsymbol\alpha^{(n+1)}}, {\boldsymbol\beta^{(n+1)}}\}, \{\mathbf{W}_t^{(n+1)}\}\big)\\
   			&\overset{(d)}\le \mathcal{R}\big(\{{\boldsymbol\alpha^{(n+1)}}, {\boldsymbol\beta^{(n+1)}}\}, \{\mathbf{W}_t^{(n+1)}\}\big),
 	\end{aligned}
 	\end{equation}
	where $\mathcal{R}^{\textup{lower}}_{\{\mathbf{W}_t^{(n)}\}}$ represents the objective function’s value of problem \eqref{P_RIS5}. (b) follows the fact that the first-order Taylor expansions are tight at the given local points in problem \eqref{P_RIS}; (c) holds since for given $\{{\boldsymbol\alpha^{(n+1)}}, {\boldsymbol\beta^{(n+1)}}\}$,
$\{\mathbf{W}_t^{(n+1)}\}$ is the optimal solution to problem \eqref{P_RIS5}; (d) is obtained since problem \eqref{P_RIS5} always provides a lower bound solution for problem \eqref{P_max}. 
  
   As a result, based on (56) and (57), we obtain that
     	\begin{align}
 		\mathcal{R}\big(\{{\boldsymbol\alpha^{(n)}}\!, {\boldsymbol\beta^{(n)}}\},\!\{\mathbf{W}_t^{(n)}\}\big)
  \!\le\!\mathcal{R}\big(\{{\boldsymbol\alpha^{(n+1)}}\!, {\boldsymbol\beta^{(n+1)}}\},\!\{\mathbf{W}_t^{(n+1)}\}\big).
   \end{align}
\begin{remark} Equation (58) shows that, the objective value of problem \eqref{P_max} is monotonically non-decreasing after each iteration of  \textbf{Algorithm 2}. On the other hand, since the achievable sum rate is upper bounded due to the restricted communication resources, a stationary point can be achieved after a finite number of iterations.
\end{remark}
\subsubsection{Complexity analysis}
The computational complexity of \textbf{Algorithm 2} can be quantified as follows.
In step 3, the complexity of exhaustively searching all possible power allocation schemes (i.e., feasible combination of $\alpha_n$ and $\beta_d$ according to (39)) with accuracy $\kappa$ is ${{\mathcal{O}}}(\frac{1}{\kappa^2})$.
In step 4, the main complexity of \textbf{Algorithm 1} is caused by solving the relaxed problem \eqref{P_RIS5} in the inner layer iteration. To solve this standard semidefinite programming (SDP) for RIS phase-shift optimization, the required complexity is ${{\mathcal{O}}}(2M^{3.5})$, if the interior-point method is employed~\cite{Luo}.
Therefore, the overall computational complexity by utilizing the proposed AO based algorithm for solving problem \eqref{P_max} for F-NOMA is ${{\mathcal{O}}}\Big(I^A_{ite}\big(\frac{1}{\kappa^2}+I_{ out}^AI_{inn}^A(2M^{3.5})\big)\Big)$, where $I_{inn}^A$ and $I_{out}^A$ respectively denote the number of iterations required for reaching convergence in the inner layer and outer layer for  \textbf{Algorithm 1}, $I^A_{ite}$ denotes the numbers of the alternating iteration for \textbf{Algorithm 2}.
\section{JO Based Approach} 
In this section, we put forward another idea to solve problem \eqref{P_max} for H-NOMA. Overcoming the optimization framework based on classical AO algorithm to alternately optimize different variables in blocks, we conceive a computationally efficient algorithm to jointly optimize the resource allocation by invoking the JO based approach.
\subsection{Joint Optimization of Resource Allocation}
Inspired by the approach introduced in Section III, we leverage the penalty based DC programming to tackle the non-convexity of rank-one constraint in RIS phase-shift configuration. The rank-relaxed problem for F-NOMA is given by
 \begin{subequations}\label{P_joint}
	\begin{align}
		&\mathop{\rm{min}}\limits_{{\boldsymbol\alpha}, {\boldsymbol\beta}, \mathbf{W}_t}\quad
		-\textit R_{sum}^{\rm{F}}+\frac{1}{\eta}  \sum_{t\in\{1,2\}}(\|\mathbf {W}_t\|_*-\overline{\mathbf {W}}^{l}_t)\\
		&\;\;\;\,{\rm s.t.} \;\;
		\eqref{alpha_max},\eqref{beta_max},\eqref{SINR_N2},\eqref{SINR_D2},	\eqref{Succeq},\eqref{SHIFT_W}.
	\end{align}
\end{subequations}
Note that, the update of the outer layer's penalty factor is in agreement with (52). Problem \eqref{P_joint} is non-convex due to the coupling of the three optimization variables (i.e. ${\boldsymbol\alpha}$, ${\boldsymbol\beta}$ and $\mathbf{W}_t$) in the inner layer SINR constraints \eqref{SINR_N2} and \eqref{SINR_D2}.

Define ${p}_{1,n}=\alpha_n P_{a}$ and ${\textit q}_{{\rm AR}}={\textup{Tr}(\mathbf {W}_1 \mathbf Z_{{\rm AR}} \mathbf Z^H_{{\rm AR}})}$, to convexify $\textup{SINR}_{r{\rightarrow}n}^{(1)}$ in constraint \eqref{SINR_N2}, which is equivalently rewritten as
	\begin{align}
	\textup{SINR}_{r{\rightarrow}n}^{(1)}=\frac{({p}_{1, n}+{\textit q}_{{\rm AR}})^2-({p}_{1, n}^2+ {q}^2_{{\rm AR}})}{2{\sigma_r^2}},
\end{align}
the key observation is that, based on the property that the first-order Taylor expansion of a convex function is a global under-estimator, $({p}_{1, n}+{q}_{{\rm AR}})^2$ can be further expanded to obtain the concave lower bound expression of $\textup{SINR}_{r{\rightarrow} n}^{(1)}$ as,
\begin{equation} 
	\begin{aligned}
		\textup{SINR}_{r{\rightarrow} n}^{(1)}
		\ge &\frac{1}{\sigma_r^2}\big({p}_{1, n}^l+ {q}^{l}_{{\rm AR}})({p}_{1, n}+{q}_{{\rm AR}})\\&-\frac{({p}_{1, n}^l+{q}^{l}_{{\rm AR}})^2}{2}-\frac{{p}_{1, n}^2+ {q}^2_{{\rm AR}}}{2}\big) \triangleq [{\mathfrak{DR}}_{r{\rightarrow} n}^{(1)}]^{\textup{lower}},
	\end{aligned}
\end{equation} 
where $\{{p}_{1, n}^l, {q}_{{\rm AR}}^l\}$ are the local points in the $l$-th iteration. 
Similarly, define ${q}_{{\rm A}n}={\textup{Tr}(\mathbf {W}_1 \mathbf Z_{{\rm A}n} \mathbf Z^H_{{\rm A}n})}$, the concave lower bound expression of $\textup{SINR}_{n{\rightarrow}n}^{(1)}$ in the $l$-th iteration is given by
\begin{equation} 
	\begin{aligned}
		\textup{SINR}_{n{\rightarrow}n}^{(1)}\ge& \frac{1}{\sigma_n^2}\big(({p}_{1, n}^l+ {q}^{l}_{{\rm A}n})({p}_{1, n}+{q}_{{\rm A}n}) \\
		&-\frac{({p}_{1, n}^l+{q}^{l}_{{\rm A}n})^2}{2}-\frac{{p}_{1, n}^2+ {q}^2_{{\rm A}n}}{2}\big) \triangleq [{\mathfrak{D}}_{n{\rightarrow} n}^{(1)}]^{\textup{lower}}.
	\end{aligned}
\end{equation} 

To achieve a more tractable concave expression of $\textup{SINR}_{n{\rightarrow} n}^{{\rm F}{(2)}}$, with reference to (49), we introduce slack variables $X_{{\rm R}n}^{'}=\frac{1}{\beta_{n}P_{r}\textup{Tr}(\mathbf {W}_2\mathbf Z_{{\rm R}n} \mathbf Z^H_{{\rm R}n})}$ and $Y_{{\rm R}n}^{'}={\beta_{d}P_{r}\textup{Tr}(\mathbf {W}_2 \mathbf Z_{{\rm R}n}\mathbf Z^H_{{\rm R}n})+\sigma_n^2}$, such that,
\begin{equation} 
	\begin{aligned}
	Y_{{\rm R}n}^{'}
	= &(1-\beta_{n})P_{r}\textup{Tr}(\mathbf {W}_2 \mathbf Z_{{\rm R}n}\mathbf Z^H_{{\rm R}n})+\sigma_n^2\\
	\le& P_{r} {q}_{{\rm R}n}-\big(({p}_{2, n}^l+ {q}^{l}_{{\rm R}n})({p}_{2, n}+ {q}_{{\rm R}n})\\
	&-\frac{({p}_{2, n}^l+ {q}^{l}_{{\rm R}n})^2}{2} -\frac{{p}_{2, n}^2+ {q}^2_{{\rm R}n}}{2}\big) +\sigma_n^2 \triangleq \Pi_{{\rm R}n},
	\end{aligned}
\end{equation} 
where ${p}_{2, n}=\beta_n P_{r}$  and $ {q}_{{\rm R}n}={\textup{Tr}(\mathbf {W}_2 \mathbf Z_{{\rm R}n} \mathbf Z^H_{{\rm R}n})}$. $\Pi_{{\rm R}n}$ is the $Y_{{\rm R}n}^{'}$ upper bound. Due to (63), we have the following transformation for  $X_{{\rm R}n}^{'}$,
\begin{equation} 
	\begin{split}
		\frac{1}{X_{{\rm R}n}^{'}}
		\ge P_r {q}_{{\rm R}n}+\sigma_n^2-\Pi_{{\rm R}n}\triangleq\Omega_{{\rm R}n},
	\end{split}
\end{equation} 
where $\Omega_{{\rm R}n}$ is the $\frac{1}{X_{{\rm R}n}^{'}}$ lower bound. Introduce the slack variable $\varphi_{{\rm R}n}$, and define a new constraint,
\begin{equation}
\Omega_{{\rm R}n}\ge\varphi_{{\rm R}n},
\end{equation}
which is further relaxed to
\begin{equation}
	X_{{\rm R}n}^{'}\le\frac{1}{\varphi_{{\rm R}n}}.
\end{equation}
Based on \textbf{Lemma 1},  the concave lower bound of $\textup{SINR}_{n{\rightarrow}n}^{{\rm F}{(2)}}$ at given local points $\{ X^{'l}_{{\rm R}n}, Y^{'l}_{{\rm R}n}\}$ can be expressed as 
\begin{equation} 
	\begin{aligned}
		\textup{SINR}_{n{\rightarrow} n}^{{\rm F}{(2)}}&\ge \frac{1}{X^{'l}_{{\rm R}n}Y^{'l}_{{\rm R}n}}-\frac{\frac{1}{\varphi_{{\rm R}n}}-X^{'l}_{{\rm R}n}}{{X^{'l}_{{\rm R}n} }^2Y^{'l}_{{\rm R}n}}-\frac{\Pi_{{\rm R}n}-Y^{'l}_{{\rm R}n}}{{Y^{'l}_{{\rm R}n} }^2X^{'l}_{{\rm R}n}}\\
		&\triangleq [{\mathfrak{D}}_{n{\rightarrow}n}^{{\rm F}{(2)}}]^{\textup{lower}}.
	\end{aligned}
\end{equation} 

As for the non-convex target SINR constraint \eqref{SINR_D2} at user $\textit d$, we can observe that the terms leading to the non-convexity of problem  \eqref{P_joint} possess the same form, which means that the similar relaxation methods in constraint \eqref{SINR_N2} are also applicable to constrain \eqref{SINR_D2}. 

For elaboration, introduce $X_{{\rm AR}}^{'}=\frac{1}{\alpha_{d}P_{a}\textup{Tr}(\mathbf {W}_1\mathbf Z_{\textup{AR } } \mathbf Z^H_{{\rm AR}})}$ and $Y_{ {\rm AR}}^{'}={\alpha_{n}P_{a}\textup{Tr}(\mathbf {W}_1 \mathbf Z_{{\rm AR}}\mathbf Z^H_{{\rm AR}})+\sigma_r^2}$ to replace $X_{{\rm R}n}^{'}$ and $Y_{{\rm R}n}^{'}$, respectively, by exploiting (67), $\textup{SINR}_{r{\rightarrow} d}^{(1)}$ can be approximated to its concave lower bound, denoted by $[{\mathfrak{D}}_{r{\rightarrow}d}^{(1)}]^{\textup{lower}}$.
In addition, $\textup{SINR}_{d{\rightarrow}d}^{(1)}$ can be coped with the same method,  where the relevant variables are further replaced by $X_{{\rm A}d}^{'}=\frac{1}{\alpha_{d}P_{a}\textup{Tr}(\mathbf {W}_1\mathbf Z_{{\rm A}d} \mathbf Z^H_{{\rm A}d})}$ and $Y_{{\rm A}d}^{'}={\alpha_{n}P_{a}\textup{Tr}(\mathbf {W}_1 \mathbf Z_{{\rm A}d}\mathbf Z^H_{{\rm A}d})+\sigma_d^2}$, and we denote the lower bound as $[{\mathfrak{D}}_{d{\rightarrow}d}^{(1)}]^{\textup{lower}}$.
Now, the remaining non-convexity is caused by $\textup{SINR}_{d{\rightarrow} d}^{{\rm F}{(2)}}$, its concave lower bound can be constructed as the expressions in (61) and (62),
\begin{equation} 
	\begin{aligned}
		\textup{SINR}_{d{\rightarrow} d}^{{\rm F}{(2)}}\ge& \frac{1}{\sigma_d^2}\big(({p}_{2, d}^l+ {q}^{l}_{{\rm R}d})({p}_{2, d}+{q}_{{\rm R}d})\\
		&-\frac{({p}_{2, d}^l+ {q}^{l}_{{\rm R}d})^2}{2}-\frac{{p}_{2, d}^2+ {q}^2_{{\rm R}d}}{2}\big) \triangleq [{\mathfrak{D}}_{d{\rightarrow} d}^{{\rm F}(2)}]^{\textup{lower}},
	\end{aligned}
\end{equation} 
where ${p}_{2, d}=\beta_d P_{r}$ and $ {q}_{{\rm R}d}={\textup{Tr}(\mathbf {W}_2 \mathbf Z_{{\rm R}d} \mathbf Z^H_{{\rm R}d})}$.

After a series of transformations to resolve the coupling between the optimization variables, problem \eqref{P_joint} for F-NOMA can be
 reformulated as
	\begin{subequations}\label{P_joint2}
	\begin{align}
		&\mathop{\rm{min}}\limits_{{\boldsymbol\alpha}, {\boldsymbol\beta}, \mathbf{W}_t}\quad
		-\textit R_{sum}^{\rm{F}}+\frac{1}{\eta}  \sum_{t\in\{1,2\}}(\|\mathbf {W}_t\|_*-\overline{\mathbf {W}}^{l}_t)\\
		& \,\;\;\; {\rm s.t.} \;\,
		 \min\Big \{
		[{\mathfrak{D}}_{r{\rightarrow} k}^{{(1)}}]^{\textup{lower}}, \big([{\mathfrak{D}}_{k{\rightarrow}k}^{(1)}]^{\textup{lower}}+[{\mathfrak{D}}_{k{\rightarrow} k}^{\rm F{(2)}}]^{\textup{lower}}\big)\Big \}\notag\\ 
		&\;\quad\quad\;\;\ge \boldsymbol{\gamma}_{k}^{\rm \min},  \forall k\in\mathcal{K},\\
		&\;\quad\quad\;\;
		\eqref{alpha_max},\eqref{beta_max},\eqref{Succeq},\eqref{SHIFT_W}.
	\end{align}
\end{subequations}
Putting all above together, problem \eqref{P_joint2} is a standard convex optimization problem and can be directly solved by the convex solver (e.g., CVX). 
\subsection{Proposed Algorithm, Convergence, and Complexity}
The details of the developed two-layer penalty based JO algorithm are summarized in \textbf{Algorithm 3}.
\begin{algorithm}[!t]\label{method1}
	\caption{Proposed Two-layer Penalty Based JO Algorithm for Solving Problem \eqref{P_max} for F-NOMA}
	\begin{algorithmic}[1]
		\STATE {{Initialize} $\big\{{\boldsymbol\alpha^{(0)}}, {\boldsymbol\beta^{(0)}},\mathbf{W}_t^{(0)}, t\in\{1, 2\}\big\}$, $\eta$, $r_1=0$, $r_2=0$,  $\epsilon $, $\varepsilon$, $r_{max}$.}
		\STATE {\bf repeat: outer loop}
		\STATE \quad Set iteration index $r_1\leftarrow0$ for inner loop.
		\STATE \quad {\bf repeat: inner loop}
		\STATE \quad\quad Solve problem \eqref{P_joint2} with given $\big\{{\boldsymbol\alpha^{(r_1)}}, {\boldsymbol\beta^{(r_1)}},\mathbf{W}_t^{(r_1)}\big\}$, and update $\big\{{\boldsymbol\alpha^{(*)}}, {\boldsymbol\beta^{(*)}},\mathbf{W}_t^{(*)}\big\}=\big\{{\boldsymbol\alpha^{(r_1+1)}}, {\boldsymbol\beta^{(r_1+1)}},\mathbf{W}_t^{(r_1+1)}\big\}$.
		\STATE \quad\quad $r_1\leftarrow r_1+1$.
		\STATE \quad {\bf until} the fractional decrease of the objective function value is below a predefined threshold $ \varepsilon >0$ or the maximum number of inner iterations $r_{\max}$ is reached.
		\STATE \quad Update $\big\{{\boldsymbol\alpha^{(0)}}, {\boldsymbol\beta^{(0)}},\mathbf{W}_t^{(0)}\big\}$ with the current solutions $\big\{{\boldsymbol\alpha^{(r_1)}}, {\boldsymbol\beta^{(r_1)}},\mathbf{W}_t^{(r_1)}\big\}$.
		\STATE \quad Update penalty factor $\eta^{(r_2)}\leftarrow c\eta^{(r_2)}$.
		\STATE \quad $r_2\leftarrow r_2+1$.
		\STATE {\bf until} the constraint violation is below a predefined threshold $\epsilon>0$.
	\end{algorithmic}
\end{algorithm}
\subsubsection{Convergence analysis}
Since both the inner-layer and the outer-layer iterations converge, the proposed algorithm always converge to a locally suboptimal solution for problem \eqref{P_max}.
\subsubsection{Complexity analysis}
The main complexity of \textbf{Algorithm 3} is caused by solving the relaxed problem \eqref{P_joint2} in the inner loop. The resulting computational complexity is 
${{\mathcal{O}}}\Big(I_{out}^JI_{inn}^J\big(2(M+2)^{3.5}\big)\Big)$ if the interior-point method is employed, where $I_{inn}^J$ and $I_{out}^J$ respectively denote the number of iterations required for reaching convergence in the inner layer and outer layer.
\section{Numerical Results}
In this section, numerical results are provided to validate the effectiveness of the coexisting passive RIS and active relay assisted NOMA system. 
\subsection{Simulation Setup}
\begin{figure}[t]
	\centering
	\setlength{\belowcaptionskip}{+0.2cm}   %调整图片标题与下文距离
	\includegraphics[width=2.8in]{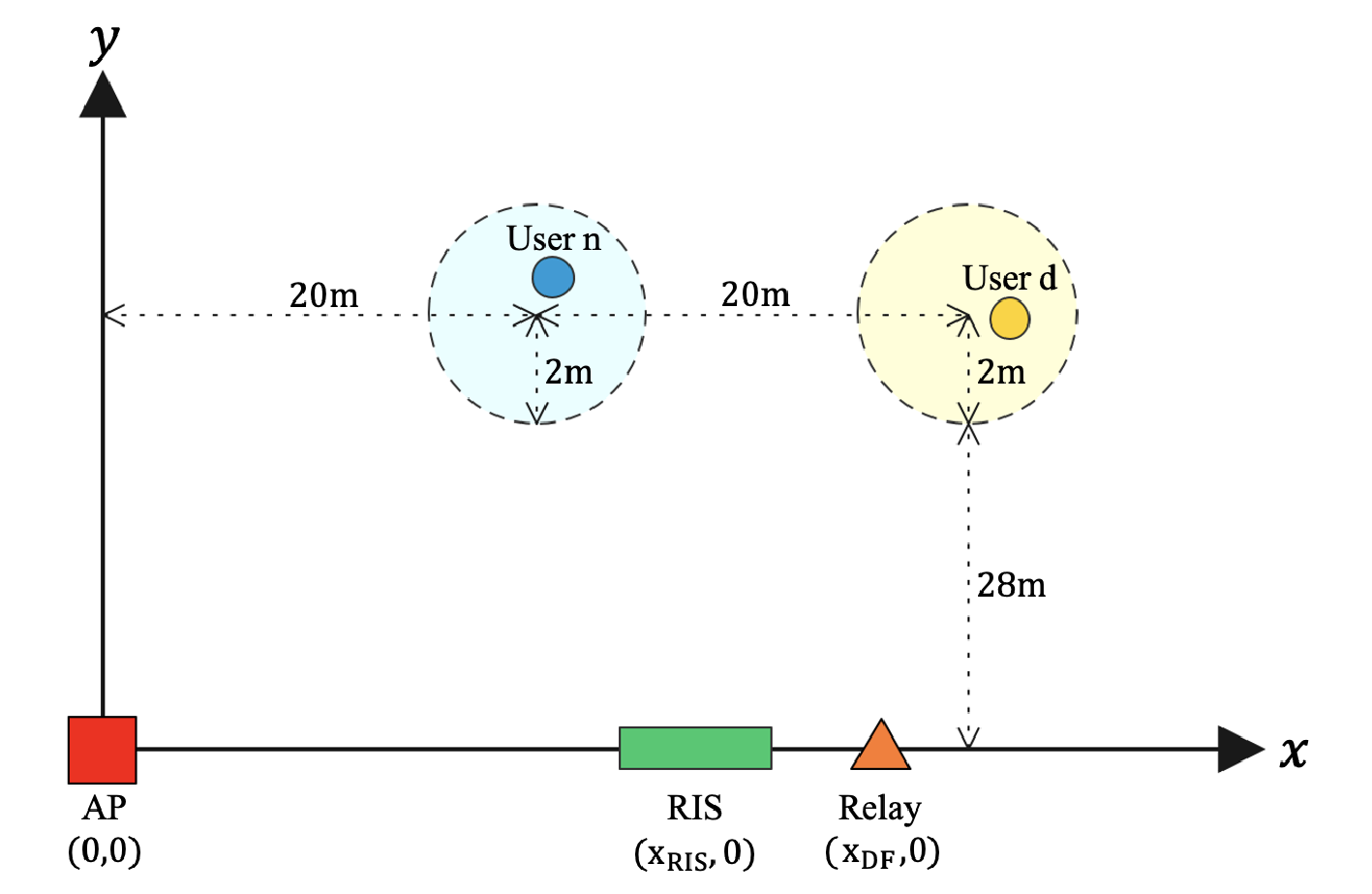}\\
	\caption{The simulated setup (top view).}\label{setup}
\end{figure}
\begin{table*}[t]%\small
	\centering
	\caption{System Parameters}
	\begin{tabular}{|l|l|}%{|c|c|}
		\hline
		\centering
		Path loss at the reference distance of 1 meter  & $\rho_0=-30 dB$ \\
		\hline
		\centering
		Rician factor of the RIS-assisted channels & $K= 3$dB \\ 
		\hline
		\centering
		Path-loss exponents of the direct channels   & $\alpha_1 = 3.5$   \\ 
		\hline
		\centering
		Path-loss exponents of the RIS-assisted channels   & $\alpha_2 = 2.2$    \\ 
		\hline
		\centering
		Noise power at receivers  & $\sigma^2 = -90$ dBm    \\ 
		\hline
		\centering
		Initialized penalty factor for Algorithms 1 and 3   & $\eta={10^{ - 4}}$   \\ 
		\hline
		\centering
		Maximum number of inner iterations of Algorithms 1 and 3  & $r_{\max}=30$   \\ 
		\hline		
		\centering
		Convergence accuracy   & ${{\varepsilon}}={10^{ -3}}, {\epsilon}={10^{ - 7}}$  \\ 
		\hline
	\end{tabular}
	\centering
	\label{Parameters}
\end{table*}
The top view of the simulation setup considered over fading channels is illustrated in Fig. 2. The RIS is located between two served users, which are randomly distributed in a circle centered at $\textup{(20\;m, 30\;m, 0\;m)}$  and $\textup{(40\;m, 30\;m, 0\;m)}$ with the radius of $2$ m, respectively. Unless otherwise stated,  the AP, RIS and relay’s locations are at $\textup{(0\;m, 0\;m, 0\;m)}$, $\textup{(32\;m, 0\;m, 1.5\;m)}$ and $\textup{(32\;m, 0\;m, 0\;m)}$ on the ground, respectively. The transmit power of the AP and relay is set as $P_T=P_a = P_r = 20$ dBm, and the paired users have a symmetric user target rate, i.e., $R_{n}^{\min}= R_{d}^{\min}=0.4$ bit/s/Hz. The distance-dependent path loss model is given by $PL(d)=\rho_0 (d)^{-\alpha}$, in which $\rho_0$ denotes the path loss at the reference distance 1m, while $d$ and $\alpha$  denote the distance and the path loss exponent between the corresponding transceiver. The Rayleigh fading channel model is modeled for the direct links. To capture both the large-scale and small-scale fading, we assume that the RIS-assisted channels are modeled as Rician fading channels. The adopted simulation parameters are presented in Table I~\cite{Deployment_Xidong}.
\subsection{Baseline Schemes}
	In order to demonstrate the benefits brought by the proposed system design, we consider the following two transmission schemes as the baseline schemes for comparison.
	\begin{itemize}
		\item \textbf{Baseline 1 (also referred to as conventional relay without RIS)}: 
		In this case, only a dedicated DF relay is deployed between the AP and two users to assist transmission, system performance can be improved by optimizing the power allocation at the transmitter.
		\item \textbf{Baseline 2 (also referred to as conventional RIS without relay)}: 
		In this case, the AP only relies on RIS to bring signal enhancement to both users. For comparative fairness, all schemes are guaranteed to achieve communication with the same total energy consumption over the operating time, i.e., the transmit power of AP in this baseline is set to be $P_a'=2P_T$. Essentially, the proposed two protocols are invalid due to the absence of DF relay. 
\end{itemize}

\begin{figure}[t]
	\centering	
	\setlength{\belowcaptionskip}{+0.2cm}   %调整图片标题与下文距离
	\subfigure[Sum rate maximization, $M=30$.]{
		\includegraphics[width=3.5in]{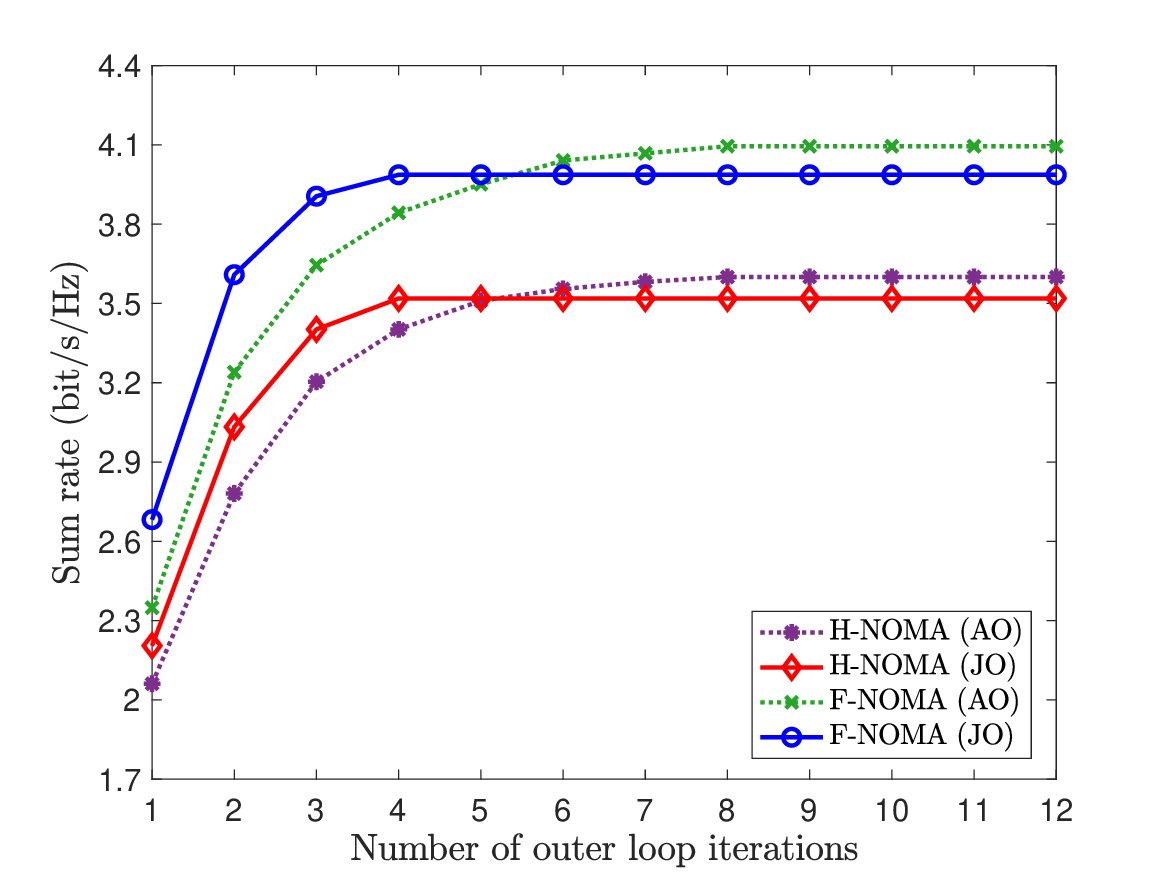}
	}
	\subfigure[Min rate maximization, $M=30$.]{
		\includegraphics[width=3.5in]{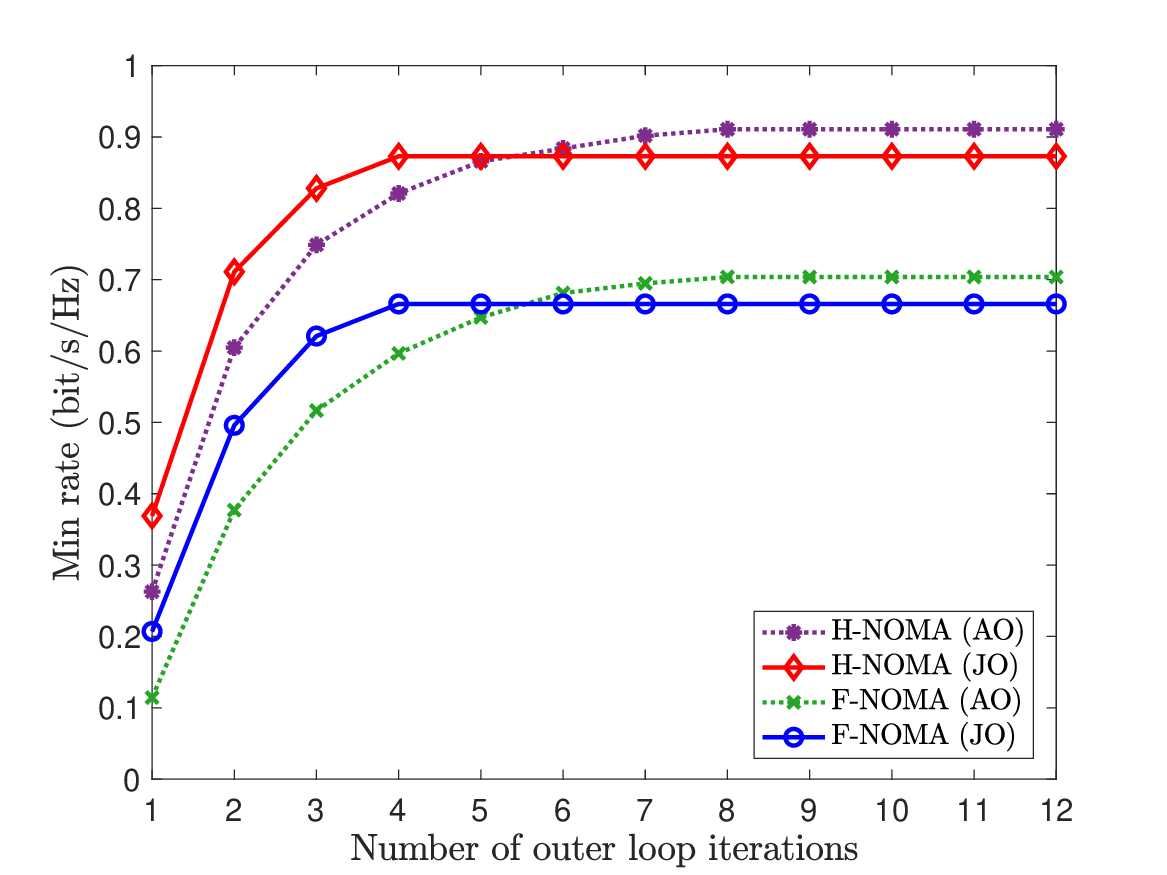}
	}
	\caption{Convergence behaviour of proposed algorithms. }
	\label{Convergence}
\end{figure}
\subsection{Convergence of the Proposed Algorithms}
Fig. 3 illustrates the convergence behavior of the proposed algorithms. For a fair comparison, with a given number of RIS reflection elements $M=30$, we compare the optimal rate value versus the number of outer loop iterations between two solutions respectively resort to AO algorithm and JO algorithm respectively. The initial power allocation coefficients $\{\boldsymbol{\alpha}, \boldsymbol {\beta}\}$ and passive beamforming vectors $\{\mathbf{v}_t, t\in\{1,2\}\}$ are obtained with the following method$\footnote{Note that, a more complex initialization scheme may further improve the convergence speed and achievable performance of the proposed algorithms, but this is beyond the scope of this paper.}$. 
	\begin{itemize}
		\item \textbf{Power allocation initialization:} 	
		For each protocol, when the communication signal covers two users at the same time, the initialization scheme of power allocation is to serve each user with equal transmit power $\frac{P_T}{2}$.	
		\item \textbf{Passive beamforming initialization:}
		Given the initial power allocation coefficients, the phase shift of each element on RIS is uniformly distributed between $[0, 2\pi)$ and the reflection amplitude is set to be 1.
\end{itemize}
As we expected, the objective function value obtained by either iterative algorithm increases rapidly with the number of outer loop iterations. Specifically, each algorithm converge within only 8 iterations and 4 iterations, respectively. For either communication protocol, JO algorithm can achieve performance approximate to AO algorithm with fewer iterations.
\subsection{Algorithm Execution Time Comparison}
\begin{figure}[t]
	\centering	
	\setlength{\belowcaptionskip}{+0.25cm}   %调整图片标题与下文距离
	\subfigure[Sum rate maximization.]{
		\includegraphics[width=3.5in]{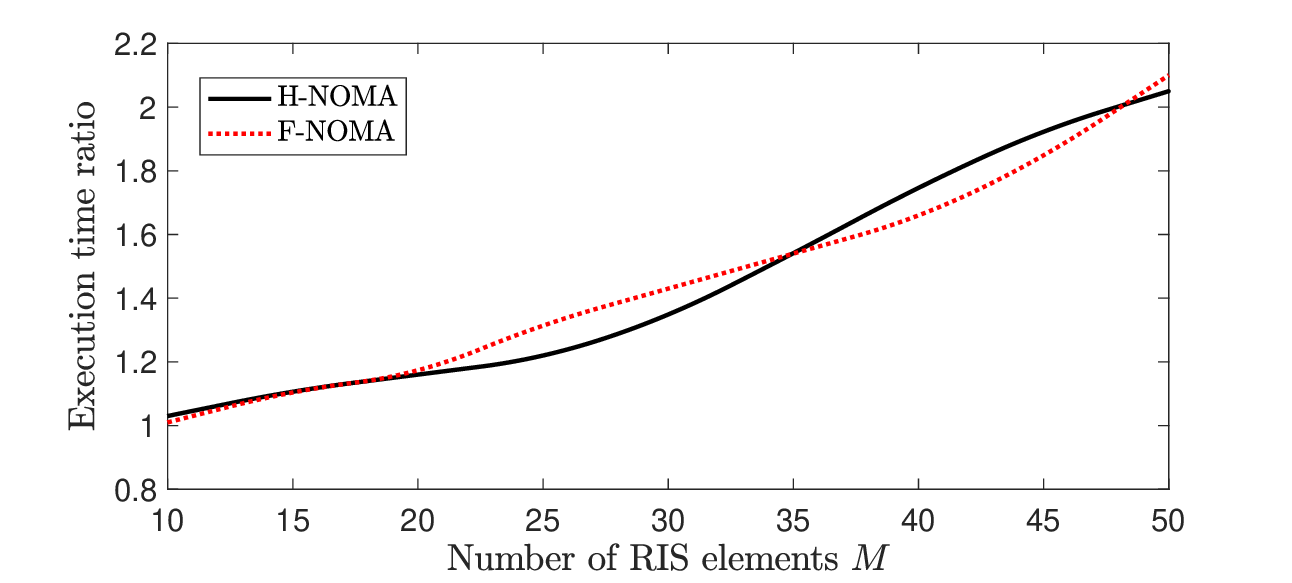}
	}
	\subfigure[Min rate maximization.]{
		\includegraphics[width=3.5in]{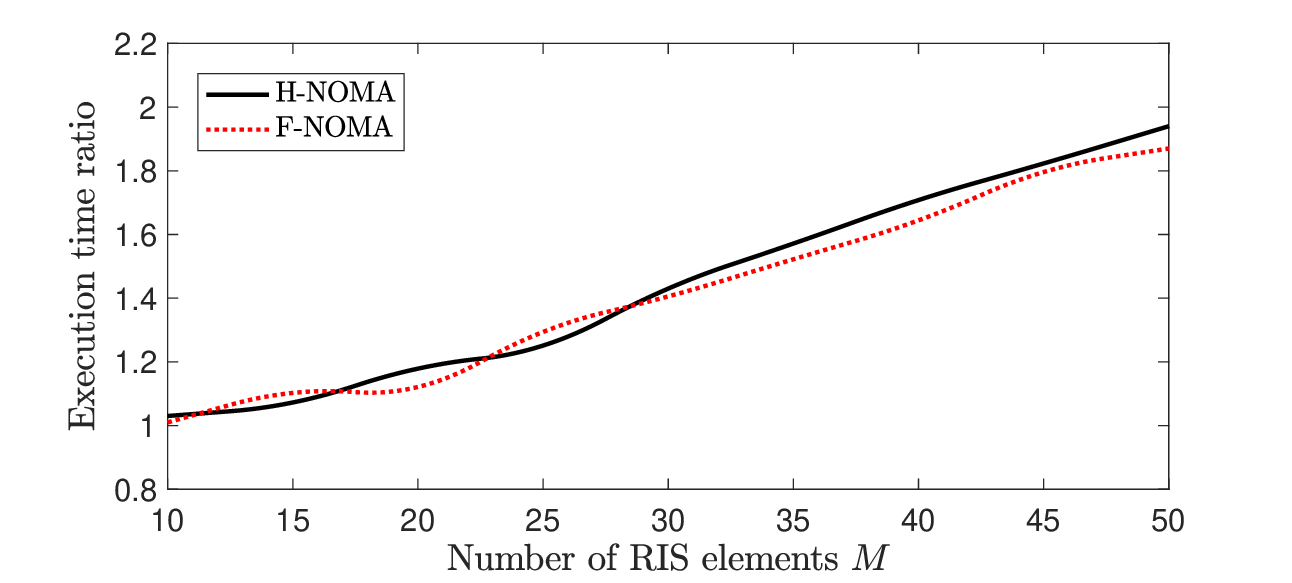}
	}
	\caption{Execution time ratio $\zeta$ between the proposed algorithms AO and JO versus the number of RIS elements $M$.}
	\label{Execution Time}
\end{figure}
Fig. 4 compares the computational complexity in terms of total execution time versus the RIS reflecting element number $M$ between proposed two algorithms, where $\zeta$ is defined as the ratio of the computational time required by algorithms AO and JO. It can be seen that the time ratio in both protocols always satisfies $\zeta>1$, which indicates that AO algorithm requires more time for convergence than JO algorithm. As the RIS reflecting element number $M$ increases, the advantage of JO over AO is observed to be more prominent. Therefore, in practice, the proposed JO algorithm is preferred since the number of RIS elements is usually large. 
\subsection{Effect of Introducing RIS in NOMA Networks}
\begin{figure}[t]
	\centering	
	\setlength{\belowcaptionskip}{+0.25cm}   %调整图片标题与下文距离
	\subfigure[Sum rate maximization.]{
		\includegraphics[width=3.5in]{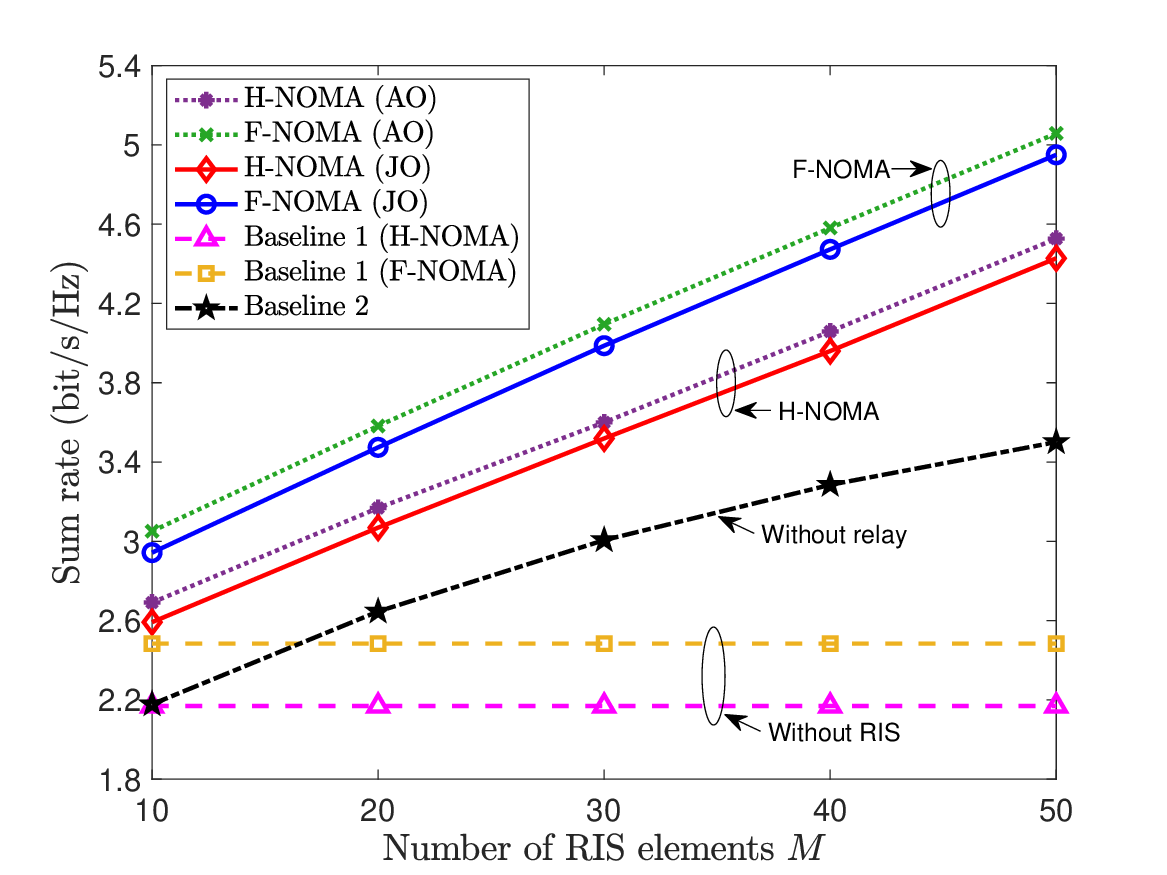}
	}
	\subfigure[Min rate maximization.]{
		\includegraphics[width=3.5in]{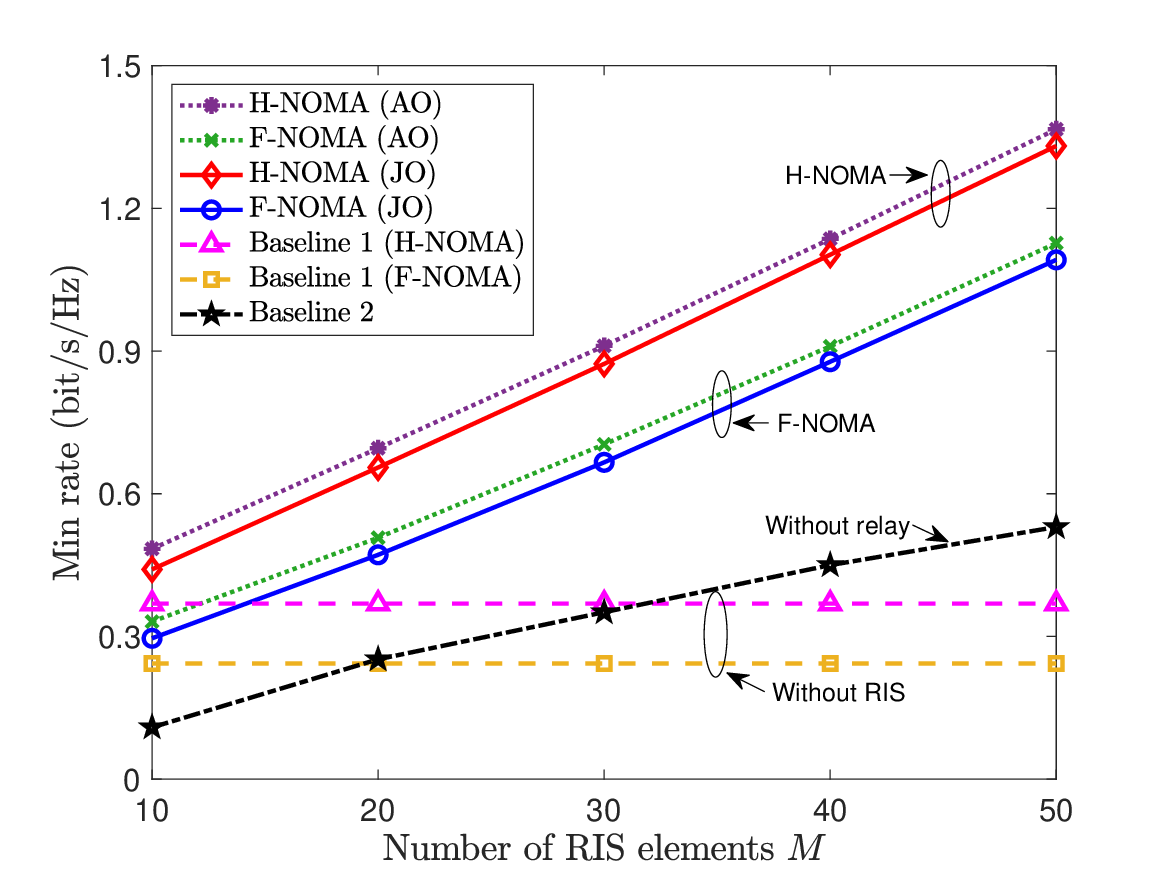}
	}
	\caption{Optimal rate versus the number of RIS elements $M$. }
	\label{RIS elements}
\end{figure}
\begin{figure}[t]
	\centering
	\setlength{\belowcaptionskip}{+0.25cm}   %调整图片标题与下文距离
	\subfigure[Sum rate maximization.]{
		\includegraphics[width=3.5in]{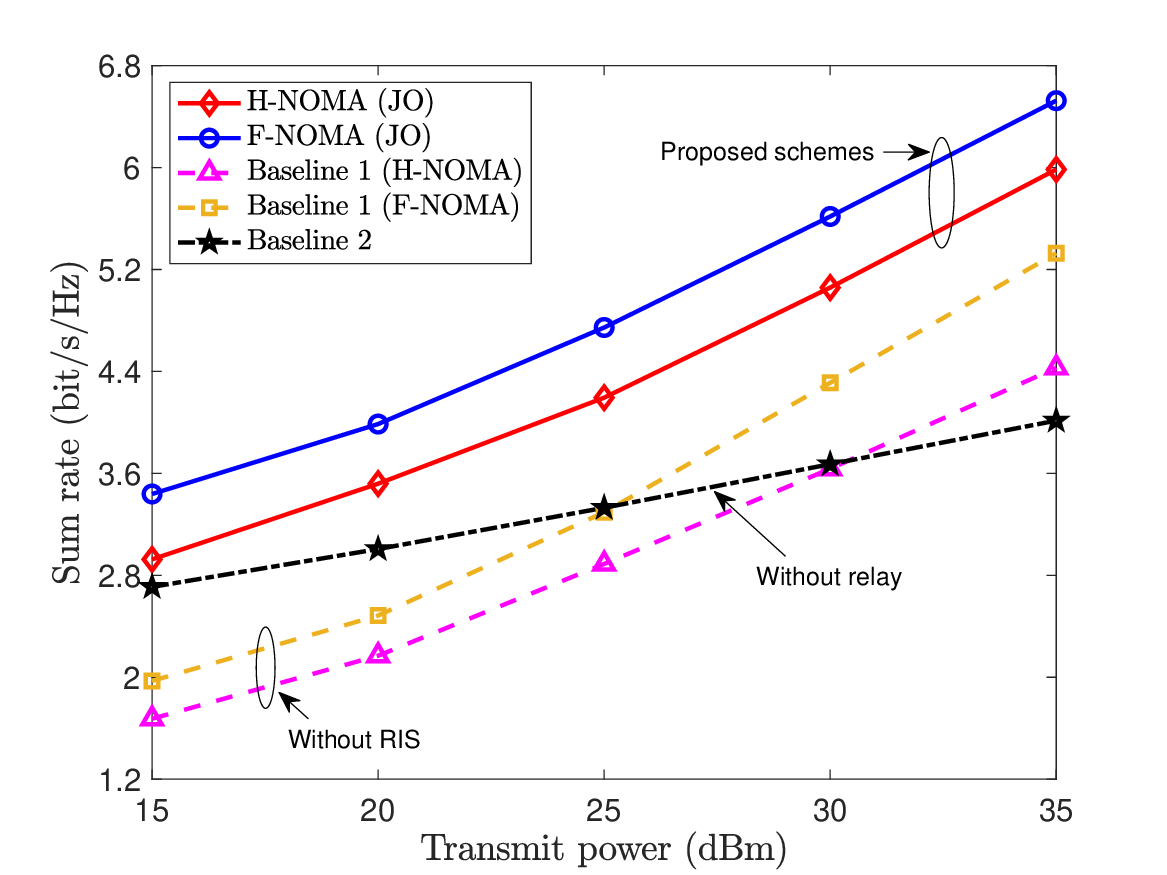}
	}
	\subfigure[Min rate maximization.]{
		\includegraphics[width=3.5in]{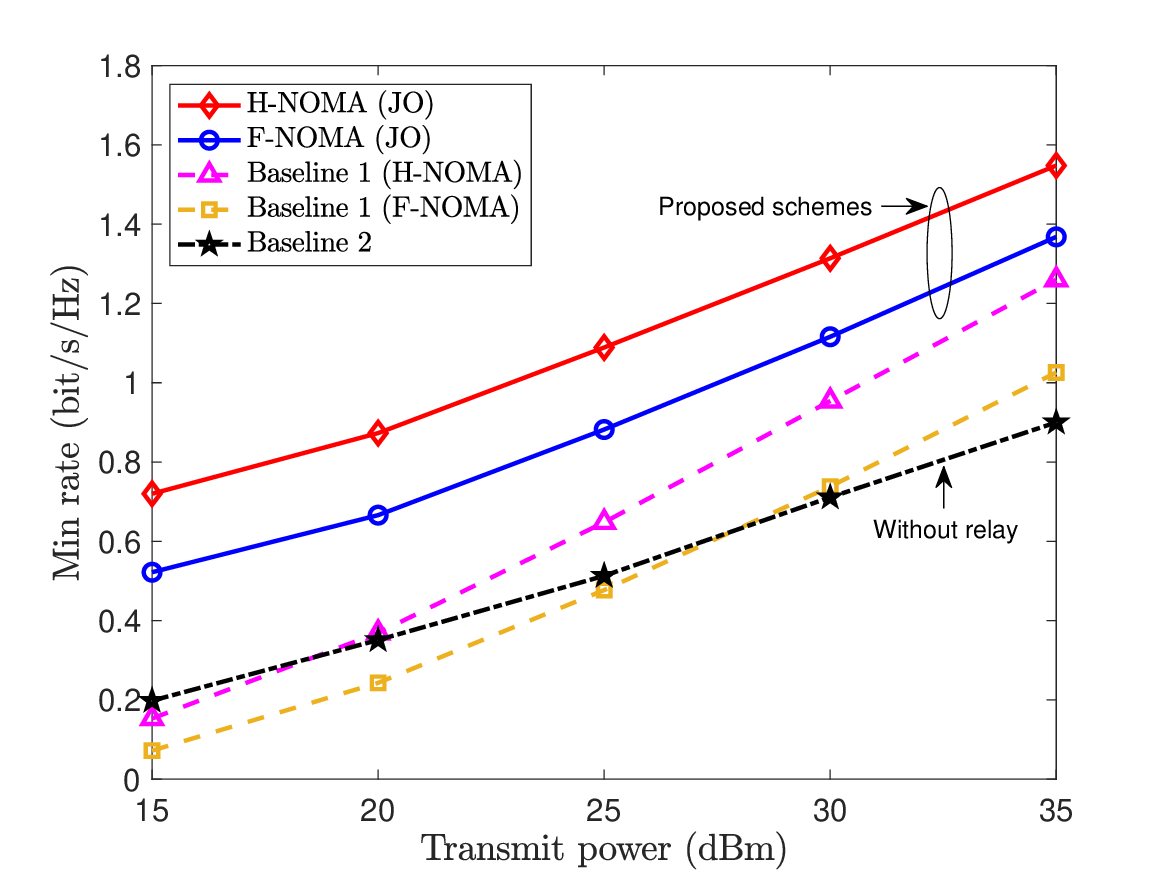}
	}
	\caption{Optimal rate versus the transmit power $P_T$ for RIS number $M=30$.}
	\label{Relay transmit power }
\end{figure}
In Fig. 5, we provide the average optimal rate versus the number of RIS reflecting elements $M$ with different schemes. It can be seen that the introduction of RIS into the system leads to a significant difference in the quality of wireless communication. Through phase-shift optimization, RIS is able to perform more efficient channel response. With the increase of reflecting elements $M$, the system performance is further enhanced. This is expected since larger arrays allow for higher gains. As for the performance of the two proposed communication protocols, F-NOMA performs better in maximizing system sum rate, whereas H-NOMA is preferable for ensuring user rate fairness. This can be explained as follows. If the relay serves both users simultaneously at IR-stage based on F-NOMA protocol, the unique degree-of-freedom provided by user scheduling supports striving for a higher achievable sum rate. In contrast, since the performance of nearby user is always stronger than that of distant user in NOMA communication, the minimum rate achievable for all links is determined by the distant user. By employing H-NOMA, distant user can fully reap the benefits of relay, which is conducive to improving user performance in the case of two-hop transmission. 
\subsection{Effect of Introducing Relay in NOMA Networks}
In Fig. 6, we present the average optimal rate versus the transmit power $P_T$ for $M=30$ with different schemes. Regarding the performance comparison among different schemes, our proposed schemes are always superior to other baseline  schemes, which benefit from the enhanced combined-channel intensity assisted by the dedicated DF relay. For achieving the same system sum rate, F-NOMA consumes less transmit power than H-NOMA. In contrast, when maximizing the minimum rate of paired users to improve user fairness, the result obtained is reversed. As can be observed from Fig. 6, there is a noticeable performance gap between both proposed schemes and the baseline scheme without relay, and this gap becomes more pronounced as the $P_T$ becomes larger. This is because higher transmitter power consumption enables more flexible synergy with RIS configuration, resulting in a more significant performance improvement.
\subsection{Effect of RIS Deployment Location }
\begin{figure}[t]
	\centering
	\setlength{\belowcaptionskip}{+0.25cm}   %调整图片标题与下文距离
	\subfigure[Sum rate maximization.]{
		\includegraphics[width=3.5in]{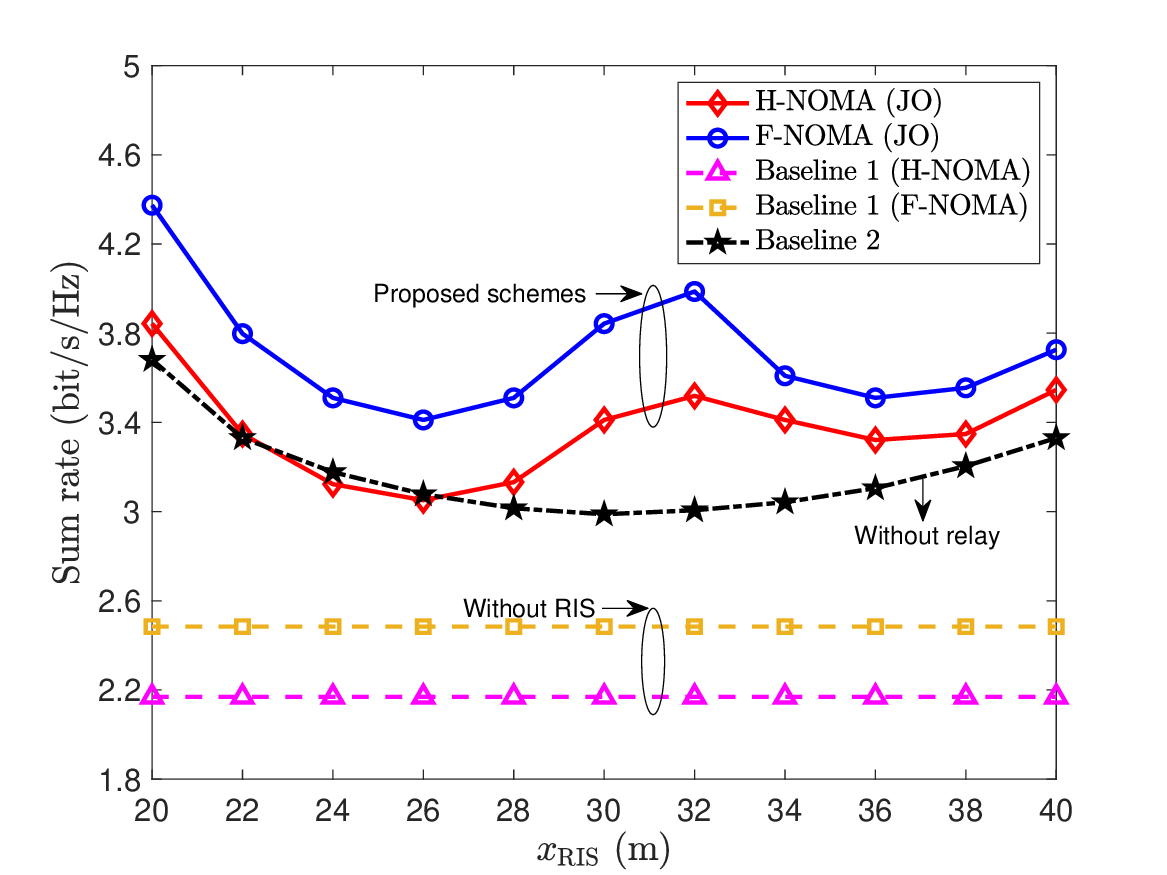}
	}
	\subfigure[Min rate maximization.]{
		\includegraphics[width=3.5in]{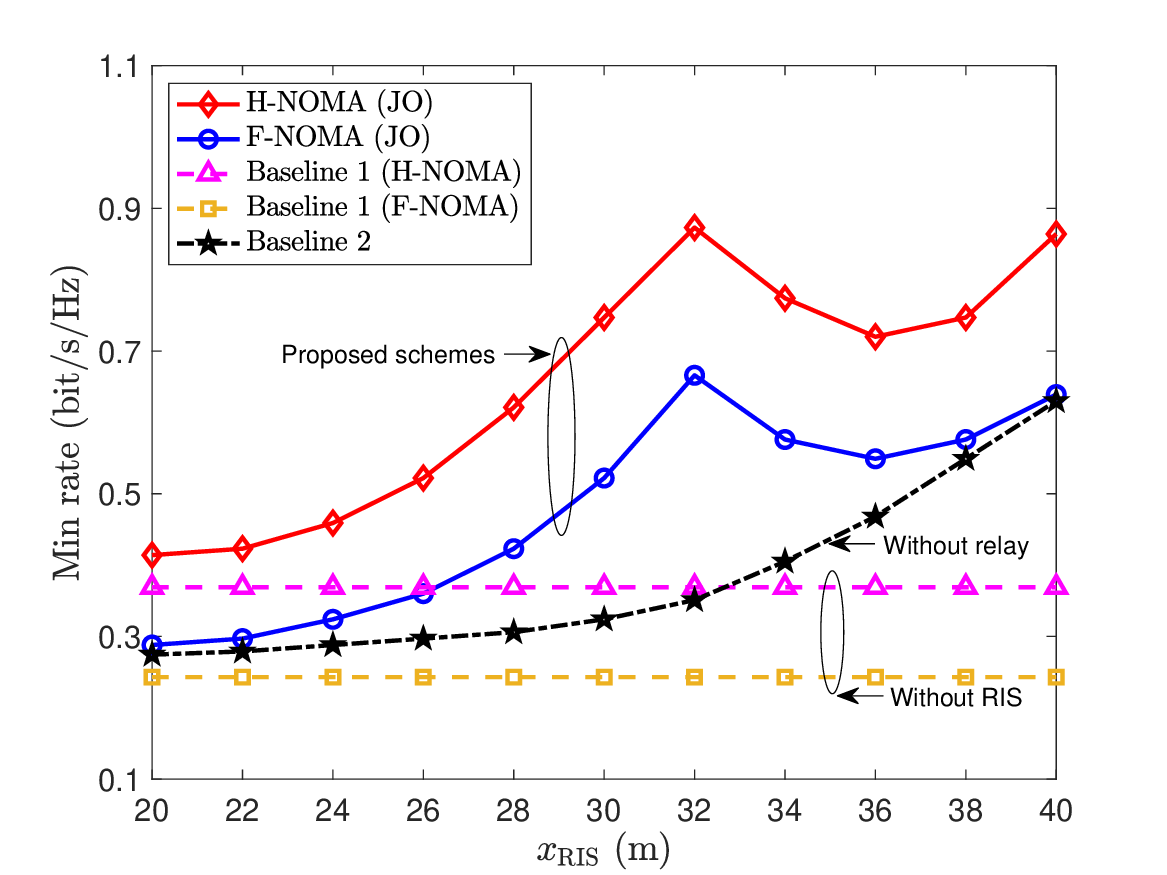}
	}
	\caption{Optimal rate versus the RIS deployment location for RIS number $M=30$. }
    \label{RIS Deployment}
\end{figure}
In Fig. 7, we investigate the impact of the RIS location on the average achievable optimal rate for RIS number $M=30$. In this paper, let $\mathcal{D}$ specify the predefined coordinates for deploying the RIS. The deployment location of the RIS should satisfy the condition $\mathcal{D}\in\{(x_{\textup{RIS}}, \textup{0\;m, 0\;m})|20\textup{\;m} \le x_{\textup{RIS}}\le 40\textup{\;m}\}$. 
To simplify the analysis, we ignore the small-scale fading effects. The system is very sensitive to the deployment location of RIS. For system sum rate maximization, it can be observed that the maximum data rate is reached when the RIS is deployed near the nearby user. It is expected that according to the characteristics of NOMA user performance distribution, the sum rate of the system is mainly contributed by the users with stronger channels, and the best communication can only be achieved when RIS serves users over a short distance. In contrast, for max-min rate optimization, system fairness is determined by the weakest user, so higher performance gains can be unlocked as RIS gets closer to the distant user. Moreover, the relay can actively forward the decoded signal resulting in a positive channel response of the system when the RIS is deployed in its vicinity.
\subsection{Effect of DF Relay Deployment Location }
Lastly, to show the impact of the DF relay deployment strategies, we depict the average optimal rate versus the relay x-axis coordinate $x_{\textup{RIS}}$ for $M=30$ in Fig. 8. The relay deployment location is predefined as $\mathcal{I}\in\{(x_{\textup{DF}}, \textup{0\;m, 0\;m})|30\textup{\;m} \le x_{\textup{DF}}\le 40\textup{\;m}\}$. It is observed that, the proposed schemes always outperform other baseline schemes. Due to the DF protocol employed, the achievable rate first increases as $x_{\textup{RIS}}$ increases until the desired maximum value is reached. Then, as the relay moves further away from this optimal deployment location, the system performance degrades accordingly, as expected.
\begin{figure}[t]
	\centering
	\setlength{\belowcaptionskip}{+0.25cm}   %调整图片标题与下文距离
	\subfigure[Sum rate maximization.]{
		\includegraphics[width=3.5in]{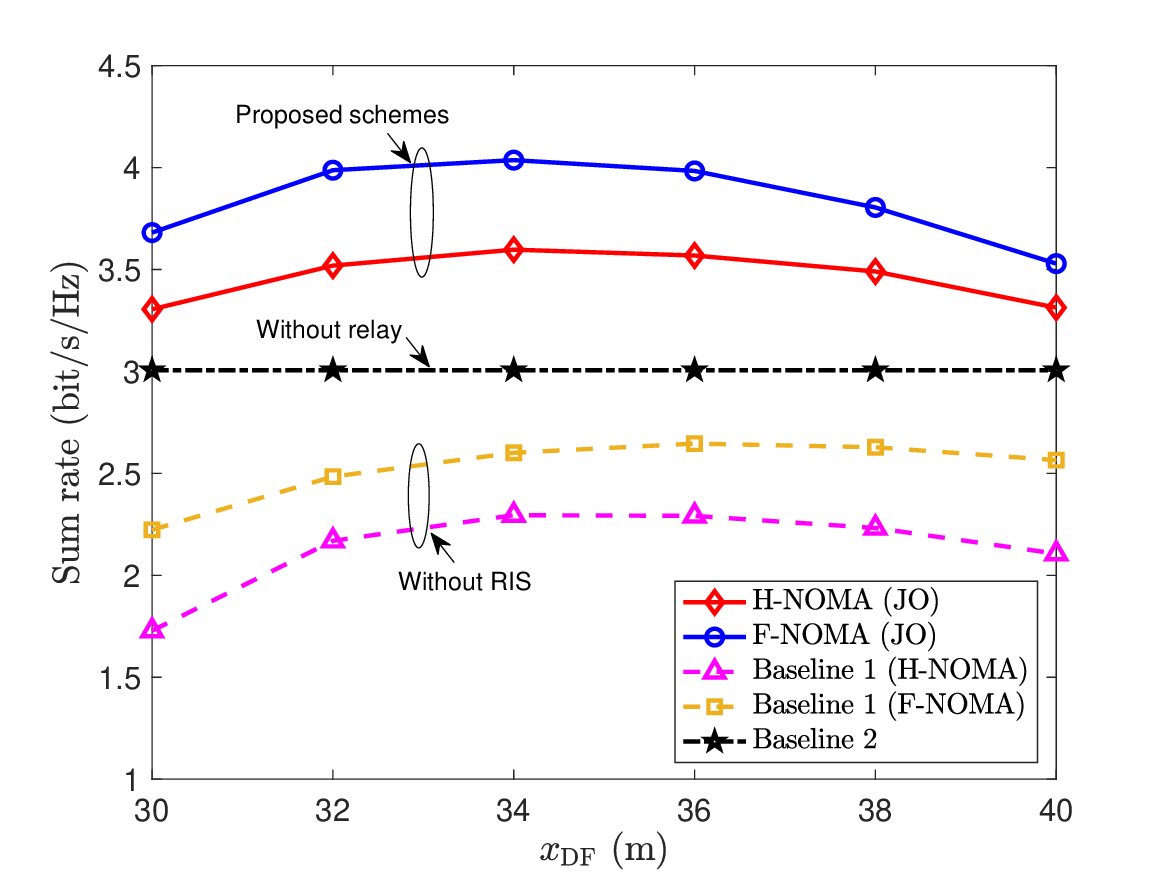}
	}
	\subfigure[Min rate maximization.]{
		\includegraphics[width=3.5in]{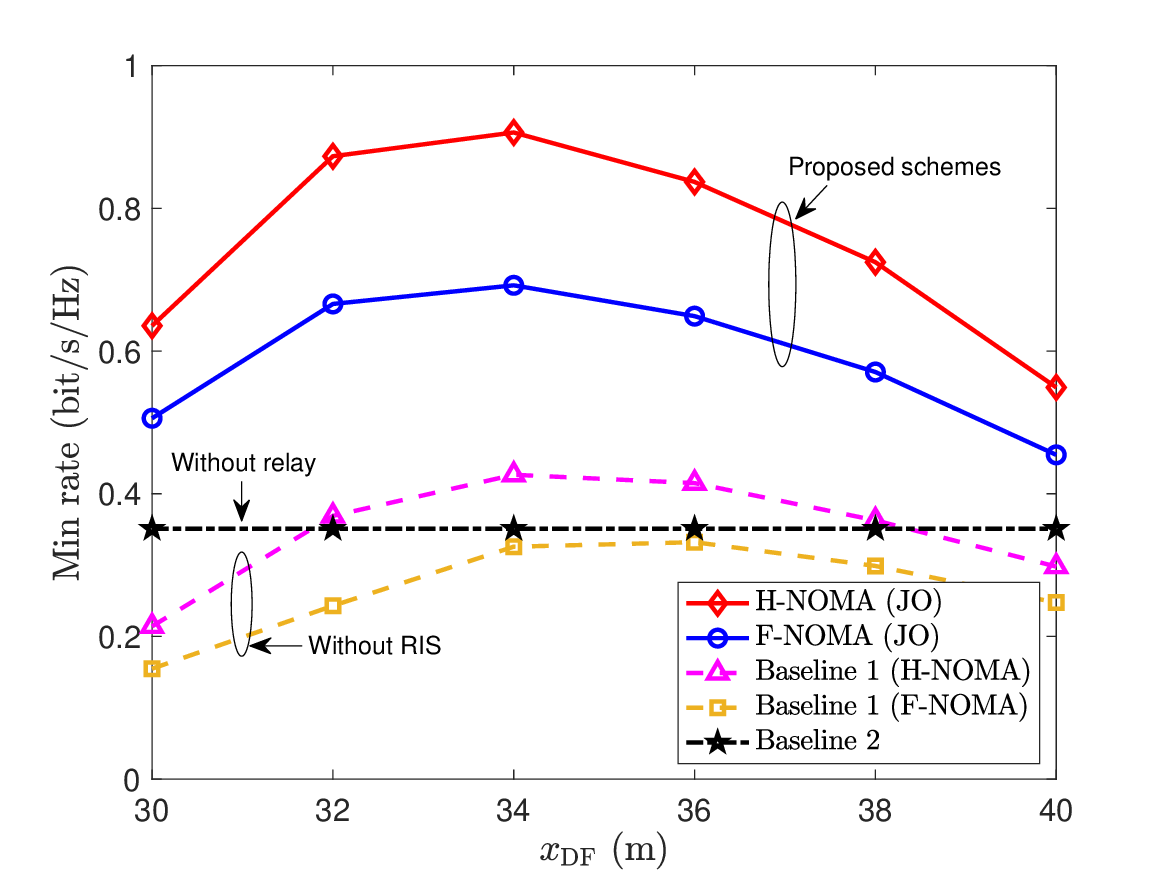}
	}
	\caption{Optimal rate versus the DF relay deployment location for RIS number $M=30$. }
	\label{Relay Deployment}
\end{figure}

\section{Conclusions}
In this paper, the coexisting passive RIS and active DF relay assisted NOMA system has been investigated. In addition, two communication protocols were proposed to pursue the potential gains of the downlink NOMA transmission framework, namely H-NOMA and F-NOMA. For each protocol, the joint resource allocation optimization problems were formulated for maximizing the system sum rate and minimum user rate. To support efficient algorithms design, the original problem was first decoupled into two subproblems and solved in an alternating manner by the proposed AO algorithm. To strike a good computational complexity-optimality trade-off, the two-layer penalty based JO algorithm was posed as a novel optimization approach to jointly optimize the resource allocation. Simulation results verified the superiority of the proposed designs. Though some performance loss is incurred by JO algorithm, its convergence speed is much faster than that of AO algorithm. Moreover, H-NOMA is preferable for ensuring the user rate fairness, while F-NOMA is a better option in terms of enhancing user sum rate. This insight provides useful guidelines for practical system implementation.

In this work, a two-stage transmission setup was considered, where the wireless resources can be fully utilized to maximize system performance gain by dynamically adjusting the resource allocation of users at different transmission stages. Despite these recent advances, extending the proposed half-duplex transmission framework to scenarios with full-duplex AP or full-duplex relay is also an important topic to be investigated. In addition, the network paradigm based on \textit{active} RIS instead of joint passive RIS and active relay can further derive additional design insights, which constitutes a promising direction for future research.
\bibliographystyle{IEEEtran}
\balance
\bibliography{mybib}

 \end{document}